# Autocorrelation-Driven Synthesis of Antenna Arrays - The Case of *DS*-Based Planar Isophoric Thinned Arrays

G. Oliveri, *Senior Member, IEEE*, G. Gottardi, *Member, IEEE*, M. A. Hannan, *Member, IEEE*, N. Anselmi, *Member, IEEE*, and L. Poli, *Member, IEEE*

***Abstract*—A new methodology for the design of isophoric thinned arrays with *a-priori* controlled pattern features is introduced. A fully-analytical and general (i.e., valid for any lattice and set of weights) relationship between the autocorrelation of the array excitations and the power pattern samples is firstly derived. Binary 2D sequences with known autocorrelation properties, namely the difference sets (*DS*s), are then chosen as a representative benchmark to prove that it is possible to deduce closed-form synthesis formulas that *a-priori* guarantee to fit requirements on the sidelobe level, the directivity, the half-power beamwidth, and the power pattern in user-defined directions. Selected results from a wide numerical assessment, which also includes full-wave simulations with realistic radiators, are illustrated to validate the reliability and the accuracy of the proposed design equations and the associated performance bounds.**

## I. Introduction and Motivations

THINNED isophoric arrays are an interesting solution in those scenario where the cost, the efficiency, the weight, the power consumption, and the number of control points of the radiating architecture are heavily constrained [1][2][3][4][5][6]. The layouts comprising equal-magnitude elements displaced in a regular grid present a higher modularity than other popular non-uniform array arrangements (e.g., sparse arrays) [7][8][9][10][11][12][13]. Unfortunately, designing isophoric thinned layouts complying with user-defined radiation constraints [e.g., bounds on the sidelobe level (*SLL*), the directivity (*D*), the half-power beamwidth ($BW^{max}$), and the null positions] is a very challenging problem because of the inherently reduced number of degrees-of-freedom (*DoF*s) for controlling the radiated power patterns [1][2]. To properly address these issues, different synthesis techniques are reported in the state-of-the-art literature (e.g., [2][3][5][6][14][15][16][17]). Early methodologies proved to be reliable in controlling the *average* sidelobes through the random removal of radiating elements from fully-populated arrangements [14][15]. More recently, the minimization of the peak sidelobe level has been yielded by means of local or global optimization-based techniques [2][3][5][6]. These latter can experience convergence issues (i.e., the convergence to sub-optimal solutions far from the optimal one) for wide apertures due to the exponential increase of the dimension of the search/solution space [2][18][19]. Otherwise, analytical thinning methods based on the exploitation of difference sets (*DS*) [16][20] and almost difference sets (*ADS*) [17][21] have been introduced. Thanks to their effectiveness and computational efficiency in dealing with arbitrary-wide isophoric arrays [16][20], these approaches have been widely adopted for solving various array synthesis problems [22][23][24][25], even though a full control of the pattern features (e.g., only the mitigation of the sidelobes) [16][17][20] is often not possible still because of the limited number of variables in designing thinned arrays.

To counteract such a limitation, this paper presents a new methodology for synthesizing thinned isophoric arrangements with controlled pattern features where the *DoF*s are (*i*) the geometry of the unit cell of the array lattice and (*ii*) the positions of the array elements on the lattice grid. Let us consider that displacing equal-magnitude elements over a non-square grid could be a promising approach to enhance the pattern control capabilities when thinning fully-populated arrangements without significantly increasing the manufacturing complexity and avoiding element tapering or sparsening. Regrettably, current thinning methodologies cannot easily handle such a design problem since (*a*) *DS*/*ADS* methods have been developed only for square lattices [16][17][20], while (*b*) applying global optimization techniques can be practically unfeasible because of the size of the search space, which is even wider than that of standard thinning problems. Therefore, a new formulation for the thinning of array layouts with *DoF*s (*i*) and (*ii*) is introduced and a set of closed-form equations are derived to synthesize the lattice geometry, the aperture size, and the thinning scheme so that the arising isophoric array *a-priori* matches user-defined pattern constraints also The authors are with the ELEDIA Research Center (ELEDIA@UniTN - University of Trento), Via Sommarive 9, Povo 38123 Trento - Italy (e-mail: {giacomo.oliveri, giorgio.gottardi, mohammadabdul.hannan, nicola.anselmi, lorenzo.poli}@unitn.it)

G. Oliveri, N. Anselmi, and L. Poli are also with the ELEDIA Research Center (ELEDIA@L2S - UMR 8506), 3 rue Joliot Curie, 91192 Gif-sur-Yvette, France (e-mail: {giacomo.oliveri, nicola.anselmi, lorenzo.poli}@l2s.centralesupelec.fr)

This work benefited from the networking activities carried out within the SNATCH Project funded by the Italian Ministry of Foreign Affairs and International Cooperation, Directorate General for Cultural and Economic Promotion and Innovation (2017-2019), the Project "WATERTECH" (Grant no. SCN_00489) funded by the Italian Ministry of Education, University, and Research (CUP: E44G14000060008), and the Project "Antenne al Plasma - Tecnologia abilitante per SATCOM (ASI.EPT.COM)" funded by the Italian Space Agency (ASI) under Grant 2018-3-HH.0 (CUP: F91I17000020005).

beyond the *SLL* generally considered in the reference literature [16][17][20]. Towards this end, first the analytic and general (i.e., not limited to a particular binary sequence such as *DS* and *ADS*) relationships between the array power pattern samples and the autocorrelation of the array excitations are derived for non-square unit cells and arbitrary array weights. Afterwards, a family of binary sequences with known autocorrelation features, namely the *DS*s, is exploited to prove the possibility to derive closed-form fully-analytical formulas for the synthesis of isophoric thinned arrays fitting user-defined pattern constraints.

The main methodological advancements with respect to the state-of-the-art lie in (*i*) the definition of general (i.e., valid for any lattice and set of excitations) relationships between the power pattern and the autocorrelation function of the array weights, (*ii*) the formulation of the thinning problem where the *DoF*s are not only the positions of the array elements on the lattice grid, but also the geometry of the unit cell of the array lattice, (*iii*) the derivation of closed-form equations for the synthesis of *DS*-based isophoric thinned arrays *a-priori* fitting pattern features also beyond the *SLL* (unlike state-of-the-art analytical thinning methods [16][17]) without requiring global search procedures (thus avoiding convergence/computational issues when handling wide apertures), and, as a byproduct, (*iv*) the generalization to arbitrary lattices of the *DS*-based thinning theory presented in [16].

The paper is organized as follows. A premise on the synthesis of arrays as the design of the autocorrelation function of a suitable excitation sequence is first done (Sect. II), then the isophoric thinned array design problem is formulated in Sect. III as a representative example of the autocorrelation-based array synthesis. Successively, a fully analytical *DS*-based design strategy is theoretically derived by determining closed-form expressions for the key figures of merit of array performance (Sect. IV). In Sect. V, a set of representative experiments, drawn from an extensive numerical validation with also realistic arrays modeled with full-wave simulations, is presented and analyzed to assess the proposed *DS*-based analytic method for isophoric thinned arrays. Some conclusions and final remarks eventually follow (Sect. VI).

## II. *Premise* - Autocorrelation-Based Array Synthesis Formulation

Let us consider a *2D* array of $P \times Q$ radiating elements located on a regular lattice characterized by a unit cell with axes $\mathbf{d}_1 = d_{1x}\hat{\mathbf{x}} + d_{1y}\hat{\mathbf{y}}$ and $\mathbf{d}_2 = d_{2x}\hat{\mathbf{x}} + d_{2y}\hat{\mathbf{y}}$ [Fig. 1(*a*)] whose radiated power pattern is given by [1]

$$\mathcal{P}(u,v) = \left| \sum_{p=0}^{P-1} \sum_{q=0}^{Q-1} E_{pq}(u,v)\, \alpha_{pq} \exp\left[ j\frac{2\pi}{\lambda} \mathbf{r}_{pq} \cdot \hat{\mathbf{r}}(u,v) \right] \right|^2 \tag{1}$$

where $E_{pq}$, $\mathbf{r}_{pq}$ ($\mathbf{r}_{pq} \triangleq p\mathbf{d}_1 + q\mathbf{d}_2$), and $\alpha_{pq}$ are the $(p,\, q)$-th ($p = 0, ..., P-1$, $q = 0, ..., Q-1$) element pattern, lattice position, and element excitation (non isophoric values - i.e., $|\alpha_{pq}| \neq 1$ - are allowed in general), respectively. Moreover, $\lambda$ is the wavelength, $\hat{\mathbf{r}}$ [$\hat{\mathbf{r}}(u,v) \triangleq (u-u_0)\hat{\mathbf{x}} + (v-v_0)\hat{\mathbf{y}}$] is the vector position being $(u,\, v)$ [$u \triangleq \sin(\theta)\cos(\varphi)$, $v \triangleq \sin(\theta)\sin(\varphi)$] and $(u_0,\, v_0)$ the cosine angles along a generic angular direction $(\theta, \varphi)$ and the steering angle $(\theta_0, \varphi_0)$.

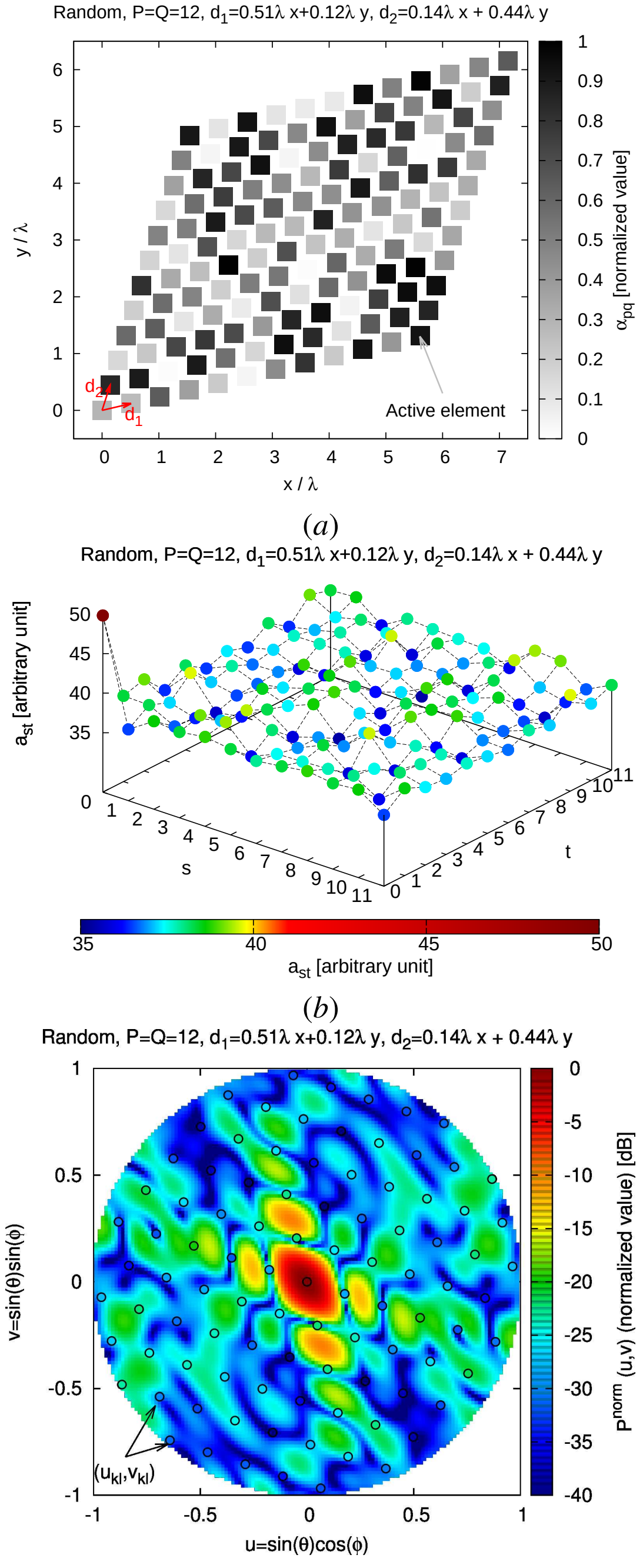

Figure 1. *Problem Geometry* - Representative example of (*a*) an array layout/excitations, (*b*) its autocorrelation function, and (*c*) the radiated pattern with samples at $(u_{kl}, v_{kl})$, $k = 0, ..., (P-1)$; $l = 0, ..., (Q-1)$ (8).

By neglecting the edge element effects and adopting the large array approximation [1], $E_{pq}(u,v) = \mathcal{E}(u,v)$, the power pattern can be rewritten as follows

$$\mathcal{P}(u,v) = \mathcal{P}_{el}(u,v) \times A(u,v) \tag{2}$$

where $\mathcal{P}_{el}(u,v)$ is the element power patter ($\mathcal{P}_{el}(u,v) \triangleq$

$|\mathcal{E}(u,v)|^2$), while

$$A(u,v) \triangleq \left|\sum_{p=0}^{P-1}\sum_{q=0}^{Q-1}\alpha_{pq}\exp\left[j\frac{2\pi}{\lambda}\mathbf{r}_{pq}\cdot\hat{\mathbf{r}}(u,v)\right]\right|^2 \quad (3)$$

is the array factor whose expression in the transformed domain $(\chi,\psi)$ ($\chi \triangleq \frac{2\pi}{\lambda}[d_{1x}(u-u_0)+d_{1y}(v-v_0)]$ and $\psi \triangleq \frac{2\pi}{\lambda}[d_{2x}(u-u_0)+d_{2y}(v-v_0)]$) is

$$A(\chi,\psi) \triangleq \left|\sum_{p=0}^{P-1}\sum_{q=0}^{Q-1}\alpha_{pq}\exp[j(p\chi+q\psi)]\right|^2 \quad (4)$$

In order to deduce some useful properties of $A(\chi,\psi)$[1], the Wiener-Khinchin theorem for Discrete Fourier Transform (*DFT*) is then exploited [27]. First, the complex function $A(\chi,\psi)$ is sampled in a set of $P$ points along $\chi$, $\underline{\chi} \triangleq \{\chi_k;\, k=0,...,(P-1)\}$, and $Q$ along $\psi$, $\underline{\psi} \triangleq \{\psi_l;\, l=0,...,(Q-1)\}$, to define the succession of $P\times Q$ samples $\underline{A}$ ($\underline{A} \triangleq \{A_{kl} \triangleq A(\chi_k,\psi_l);\; k=0,...,(P-1);\; l=0,...,(Q-1)\}$). Towards this end, it is profitable to notice that $A(\chi,\psi)$ is a periodic function with period $2\pi$, thus the sampling locations can be chosen in the ranges $\chi\in[0,2\pi]$ and $\psi\in[0,2\pi]$ as follows $\chi_k \triangleq \frac{2\pi}{P}k$ $[k=0,...,(P-1)]$ and $\psi_l \triangleq \frac{2\pi}{Q}l$ $[l=0,...,(Q-1)]$.

The sequence $\underline{A}$ of $P\times Q$ complex samples can be represented as the $DFT$ of another $P\times Q$ complex sequence $\underline{a}$ ($\underline{a} \triangleq \{a_{st};\, s=0,...,(P-1);\, t=0,...,(Q-1)\}$), $\underline{A} \triangleq \mathcal{F}\{\underline{a}\}$, its $(k,l)$-th entry being defined as $A_{kl} \triangleq \sum_{s=0}^{P-1}\sum_{t=0}^{Q-1}a_{st}\exp\left[-j\left(\frac{2\pi s}{P}k+\frac{2\pi t}{Q}l\right)\right]$. On the other hand, it can be proved (Appendix A.1) that the sequence $\underline{a}$ is equal to the discrete autocorrelation function of the sequence of the array excitations $\underline{\alpha}$ ($\underline{\alpha} \triangleq \{\alpha_{pq};\, p=0,...,(P-1);\, q=0,...,(Q-1)\}$), namely

$$\begin{array}{c} a_{st} = \sum_{p=0}^{P-1}\sum_{q=0}^{Q-1}\alpha_{pq}\alpha_{\lfloor p+s\rfloor_P\lfloor q+t\rfloor_Q} \\ s=0,...,P-1;\; t=0,...,Q-1 \end{array} \quad (5)$$

($\lfloor\cdot\rfloor$ being the "modulo" operator). Therefore, it turns out that $\underline{A}=\mathcal{F}\{\underline{a}\}$, that is

$$A_{kl} \triangleq A(\chi_k,\psi_l) = \xi_{kl} \quad (6)$$

$[k=0,...,(P-1);\, l=0,...,(Q-1)]$ being

$$\xi_{kl} \triangleq \sum_{s=0}^{P-1}\sum_{t=0}^{Q-1}a_{st}\exp\left[j\left(\frac{2\pi s}{P}k+\frac{2\pi t}{Q}l\right)\right]. \quad (7)$$

Since $\mathcal{P}(\chi_k,\psi_l) = \mathcal{P}_{el}(\chi_k,\psi_l)\times A_{kl}$, it is then enough to come back from the $(\chi_k,\psi_l)$-domain to the $(u_{kl},v_{kl})$ by applying the following transformations

$$\begin{array}{l} u_{kl} = u_0 + \frac{\lambda}{PQ}\frac{kQd_{2y}-lPd_{1y}}{\nu} \\ v_{kl} = v_0 + \frac{\lambda}{PQ}\frac{lPd_{1x}-kQd_{2x}}{\nu} \end{array} \quad (8)$$

($\nu \triangleq d_{1x}d_{2y}-d_{2x}d_{1y}$) and using (6) to predict the samples of the power pattern $\mathcal{P}(u_{kl},v_{kl})$ $[k=0,...,(P-1);$ $l=0,...,(Q-1)]$ from the knowledge of the samples of the element power pattern, $\mathcal{P}_{el}(u_{kl},v_{kl})$, and of the autocorrelation function, $\underline{a}$, of the excitation sequence, $\underline{\alpha}$,

$$\mathcal{P}(u_{kl},v_{kl}) = \mathcal{P}_{el}(u_{kl},v_{kl})\times\xi_{kl} \quad (9)$$

or in compact form

$$\underline{\mathcal{P}} = \underline{\mathcal{P}}_{el}\times\mathcal{F}\{\underline{a}\} \quad (10)$$

where $\underline{\mathcal{P}} \triangleq \{\mathcal{P}_{kl} \triangleq \mathcal{P}(u_{kl},v_{kl});\; k=0,...,(P-1);\; l=0,...,(Q-1)\}$, $\underline{\mathcal{P}}_{el} \triangleq \{\mathcal{P}_{el}(u_{kl},v_{kl});\, k=0,...,(P-1);\, l=0,...,(Q-1)\}$, and $\mathcal{F}\{\underline{a}\} \triangleq \{\xi_{kl} = \sum_{s=0}^{P-1}\sum_{t=0}^{Q-1} a_{st}\exp\left[j\left(\frac{2\pi s}{P}k+\frac{2\pi t}{Q}l\right)\right];\; k=0,...,(P-1);\; l=0,...,(Q-1)\}$.

Finally, it is also possible to yield the power pattern for every angular coordinate $(u,v)$ thanks to the interpolation formula for *DFT*. Indeed, the function $A(\chi,\psi)$ can be determined from its samples $A_{kl} \triangleq A(\chi_k,\psi_l)$ as follows [16][17][27]

$$\begin{array}{l} A(\chi,\psi) = \\ = \left|\sum_{k=0}^{P-1}\sum_{l=0}^{Q-1}\sqrt{A(\chi_k,\psi_l)}\exp[j\eta_{kl}]\,\mathcal{S}(\chi-\chi_k,\psi-\psi_l)\right|^2 \end{array} \quad (11)$$

$\eta_{kl}$ being a deterministic phase term equal to the phase of the $(k,l)$-th term of the *DFT* of the excitation sequence [17], $\underline{\eta}$ = phase$\{\mathcal{F}\{\underline{\alpha}\}\}$ ($\underline{\eta} = \{\eta_{kl} \triangleq$ phase $\{\sum_{p=0}^{P-1}\sum_{q=0}^{Q-1}\alpha_{pq}\exp\left[j\left(\frac{2\pi s}{P}k+\frac{2\pi t}{Q}l\right)\right]\};\; k=0,...,(P-1);\; l=0,...,(Q-1)\}$), where

$$\begin{array}{r} \mathcal{S}(\chi,\psi) = \frac{\sin(\chi P/2)}{P\sin(\chi/2)}\frac{\sin(\psi Q/2)}{Q\sin(\psi/2)}\times \\ \times\exp\left[j\left(\chi\frac{P-1}{2}+\psi\frac{Q-1}{2}\right)\right] \end{array} \quad (12)$$

is the interpolation function[2]. Moreover, since $A_{kl}=\xi_{kl}$ then

$$A(\chi,\psi) = \left|\sum_{k=0}^{P-1}\sum_{l=0}^{Q-1}\sqrt{\xi_{kl}}\exp[j\eta_{kl}]\,\mathcal{S}(\chi-\chi_k,\psi-\psi_l)\right|^2, \quad (13)$$

and $\mathcal{P}(u,v)$ can be easily determined by substituting (13) in (2), while using the definition of the sampling coordinates $\chi_k$ and $\psi_l$ as well as the expression of $\chi$ and $\psi$ in terms of the angular coordinates $(u,v)$

$$\begin{array}{l} \mathcal{P}(u,v) = \mathcal{P}_{el}(u,v)\times\left|\sum_{k=0}^{P-1}\sum_{l=0}^{Q-1}\left[\sqrt{\xi_{kl}}\exp(j\eta_{kl})\right]\right. \\ \quad\times\mathcal{S}\left(\frac{2\pi}{\lambda}[d_{1x}(u-u_0)+d_{1y}(v-v_0)]-\frac{2\pi k}{P},\right. \\ \quad\left.\left.\frac{2\pi}{\lambda}[d_{2x}(u-u_0)+d_{2y}(v-v_0)]-\frac{2\pi l}{Q}\right)\right|^2 \end{array} \quad (14)$$

It is worth pointing out that the expressions (9) and (14) hold true for any $\underline{\alpha}$ excitation sequence (i.e., binary excitations → "thinned arrays", zero/non-zero complex excitations → "sparse arrays", non-zero complex excitations → "fully-populated arrays", clustered excitations →"clustered arrays", etc..) and any array lattice. Moreover, they can be seen as

[1]It is worth remarking that, since $A(\chi,\psi)$ can be seen as a power density function [16], similar conclusions can be drawn by using the autocorrelation theorem and computing the Fourier transform of the autocorrelation of the *location function* associated to $\alpha_{pq}$, $p=0,...,P-1$, $q=0,...,Q-1$ [26]. Moreover, no multiplication factors are considered in (4) since the derivation refers to normalized quantities [26].

[2]The relation in (11) coincides with (9) when $(\chi,\psi)=(\chi_k,\psi_l)$, since $|\exp[j\eta_{kl}]|^2=1$ and $\mathcal{S}(\chi_k-\chi_{k'},\psi_l-\psi_{l'})=1$ if $(k',l')=(k,l)$, while $\mathcal{S}(\chi_k-\chi_{k'},\psi_l-\psi_{l'})=0$ if $(k',l')\neq(k,l)$ [27].

the "premises" to a new methodological approach to the array synthesis since they enable to

- exactly predict the samples of $\mathcal{P}(u,v)$ at (8) [Fig. 1(*c*)], without requiring the knowledge of the excitation sequence, $\underline{\alpha}$, but only the values of its autocorrelation function [Fig. 1(*b*)] and the samples of the element factor

$$\begin{array}{c}\mathcal{P}(u_{kl},v_{kl}) = \mathcal{P}_{el}(u_{kl},v_{kl}) \times \\ \times \sum_{s=0}^{P-1}\sum_{t=0}^{Q-1} a_{st} \exp\left[j\left(\frac{2\pi s}{P}k + \frac{2\pi t}{Q}l\right)\right];\end{array} \quad (15)$$

- faithfully determine the power pattern $\mathcal{P}(u,v)$ in the whole $(u,v)$-space from the knowledge of the excitation sequence, $\underline{\alpha}$, starting from the pattern samples (15), $\{\mathcal{P}(u_{kl},v_{kl});\ k = 0,...,(P-1);\ l = 0,...,(Q-1)\}$, the computation of the *DFT* of the excitation sequence, $\underline{\eta} = \text{phase}\{\mathcal{F}\{\underline{\alpha}\}\}$, and the exploitation of the interpolation function (12)

$$\begin{array}{c}\mathcal{P}(u,v) = \mathcal{P}_{el}(u,v) \times \Big| \sum_{k=0}^{P-1}\sum_{l=0}^{Q-1} \\ \Big[\sqrt{\sum_{s=0}^{P-1}\sum_{t=0}^{Q-1}\sum_{p=0}^{P-1}\sum_{q=0}^{Q-1} \alpha_{pq}\alpha_{\lfloor p+s\rfloor_P \lfloor q+t\rfloor_Q}} \\ \sqrt{\exp\left[j\left(\frac{2\pi s}{P}k + \frac{2\pi t}{Q}l\right)\right]}\exp(j\eta_{kl})\Big] \\ \times\mathcal{S}\left(\frac{2\pi}{\lambda}\left[d_{1x}(u-u_0) + d_{1y}(v-v_0)\right] - \frac{2\pi k}{P},\right. \\ \left.\frac{2\pi}{\lambda}\left[d_{2x}(u-u_0) + d_{2y}(v-v_0)\right] - \frac{2\pi l}{Q}\right)\Big|^2\end{array} \quad (16)$$

- approximate the power pattern $\mathcal{P}(u,v)$ in the whole $(u,v)$-space from the knowledge of the autocorrelation function $\underline{a}$ of the excitation sequence, $\underline{\alpha}$, and the exploitation of the interpolation function (12)

$$\begin{array}{c}\mathcal{P}(u,v) \approx \mathcal{P}_{el}(u,v) \times \Big| \sum_{k=0}^{P-1}\sum_{l=0}^{Q-1} \\ \Big[\sqrt{\sum_{s=0}^{P-1}\sum_{t=0}^{Q-1} a_{st}\exp\left[j\left(\frac{2\pi s}{P}k + \frac{2\pi t}{Q}l\right)\right]} \\ \mathcal{C}\{\exp(j\eta_{kl})\}\Big] \times \mathcal{S}\left(\frac{2\pi[d_{1x}(u-u_0)+d_{1y}(v-v_0)]}{\lambda} - \frac{2\pi k}{P},\right. \\ \left.\frac{2\pi[d_{2x}(u-u_0)+d_{2y}(v-v_0)]}{\lambda} - \frac{2\pi l}{Q}\right)\Big|^2\end{array} \quad (17)$$

$\mathcal{C}\{\,.\,\}$ being the Monte Carlo random estimation [16][17].

## III. Autocorrelation-Based *2D* Isophoric Thinning

Because of the generality of the expressions (9) and (14) and to show the potentialities of an autocorrelation-based approach to the array synthesis, let us consider as a representative benchmark the design of two-dimensional isophoric thinned arrays. In this case, the excitation sequence $\underline{\alpha}$ is a binary sequence (i.e., $\alpha_{pq} \in \{0, 1\}$, $p = 0,...,P-1$, $q = 0,...,Q-1$) that can be interpreted as a *location function* [16] and the original thinning problem "*to find the aperture size* $P \times Q$*, the unit cell axes* $\mathbf{d}_1$ *and* $\mathbf{d}_2$*, and the binary sequence* $\underline{\alpha}$ *so that the figures of merit of the corresponding radiated pattern* $\mathcal{P}(u,v)$ *satisfy user-defined constraints*" can be then re-formulated as follows "*to find the aperture size* $P \times Q$*, the unit cell axes* $\mathbf{d}_1$ *and* $\mathbf{d}_2$*, and the autocorrelation function* $\underline{a}$ *so that the figures of merit of the corresponding radiated pattern* $\mathcal{P}(u,v)$ *satisfy user-defined constraints*". As a matter of fact, the relation (15) states that the samples of $\mathcal{P}(u,v)$ only depend on the values of the autocorrelation function of $\underline{\alpha}$ (not directly on $\underline{\alpha}$) [Fig. 1(*b*)] and the samples of the element factor. Thus, the "control" of the locations/values of the pattern samples [circles in Fig. 1(*c*)], by suitably choosing/setting the autocorrelation function, allows one to constraint the figures of merit of the corresponding pattern. Concerning the design of the autocorrelation function there are two (or even more) possible approaches: (*1*) synthesizing a sequence $\underline{\alpha}$ by means an optimization procedure so that the corresponding autocorrelation fits a user-defined autocorrelation-mask or (*2*) exploiting state-of-the-art binary sequences with known autocorrelation function having suitable properties. This latter is the case of analytical *2D DS*s-based sequences [16] used in the following.

## IV. *DS*-Based Analytic *2D* Isophoric Thinning

The use of *DS* sequences [16] to address the *Isophoric Thinning Problem* is motivated by the following reasons: (*i*) *DS*s are natural candidates for array thinning since the associated sequences are binary [16], (*ii*) the cyclic autocorrelation of the *DS*-based excitations is *a-priori* known and the corresponding *DFT* can be computed in closed form [16][17], (*iii*) a wide number of *DS*s is available in open repositories [28] and additional sequences can be constructed by means of *ad-hoc* theorems [29] or synthesis techniques [30], and (*iv*) several trade-off thinned solutions can be generated from a single *DS* by using the cyclic shift property [16][17]. On the other hand, it is worthwhile to point out that the exploitation of *DS* sequences in this paper turns out to be a significant generalization of the problem addressed in [16] since, besides extending the constrained synthesis to other - and different from the *SLL* - figures of merit, additional *DoF*s, such as $\mathbf{d}_1$ and $\mathbf{d}_2$, are handled.

Let us now summarize the definition and the properties of *DS* sequences useful for the following. A $(P \times Q, H, \gamma)$-*DS*, $\underline{\Xi}$, is a set of $H$ pairs of integers whose associated binary sequence

$$\begin{array}{c}\underline{\alpha} \triangleq \{\alpha_{pq} = 1\,(0)\text{ if }(p,q) \in (\notin)\,\Xi; \\ p = 0,...,P-1,\ q = 0,...,Q-1\}\end{array} \quad (18)$$

has a two-level autocorrelation [16]

$$a_{st} = (H-\gamma)\,\zeta_{st} + \gamma \quad (19)$$

$H$ being the number of active elements in the $P \times Q$ aperture, while $\zeta_{st} = 1$ if $s = t = 0$ and $\zeta_{st} = 0$ otherwise [$s = 0,...,(P-1)$; $t = 0,...,(Q-1)$]. Owing to the properties of *DS*s [16][17], (19) holds also true when $\underline{\alpha}$ is replaced by its shifted version

$$\begin{array}{r}\underline{\alpha}^{(\sigma_x,\sigma_y)} = \left\{\alpha_{pq} = 1\,(0)\text{ if }\left(\lfloor p+\sigma_x\rfloor_P\,,\,\lfloor q+\sigma_y\rfloor_Q\right) \in (\notin)\,\Xi;\right. \\ \left. p = 0,...,P-1,\ q = 0,...,Q-1\right\}\end{array} \quad (20)$$

$\sigma_x$ and $\sigma_y$ being the integer cyclic shifts applied along $x$ and $y$, respectively, to the planar binary sequence $\underline{\alpha}$.[3]

By substituting (19) in (6), it turns out that

$$\xi_{kl} = \gamma\,(PQ\zeta_{kl} - 1) + H \quad (21)$$

[3]For the sake of notation simplicity, the shift superscript will be neglected in the following and, unless explicitly stated, all the incoming results will be considered valid for any $(\sigma_x,\sigma_y)$.

$[k = 0, ..., (P-1);\ l = 0, ..., (Q-1)]$ is a discrete function always constant to the value $\xi_{kl} = H - \gamma$ ($\forall\ k \times l \neq 0$) unless a peak equal to $\xi_{kl} = \gamma \times (P \times Q - 1) + H$ at $k = l = 0$. Moreover, the power pattern of a *DS*-based isophoric array with arbitrary unit cells can be obtained from (14) by setting $\xi_{kl}$ as in (21)

$$\mathcal{P}(u,v) = \mathcal{P}_{el}(u,v) \times \left|\sum_{k=0}^{P-1}\sum_{l=0}^{Q-1}\left[\sqrt{\gamma\left(PQ\zeta_{kl}-1\right)+H}\exp\left(j\eta_{kl}\right)\right] \times \mathcal{S}\left(\frac{2\pi}{\lambda}\left[d_{1x}(u-u_0)+d_{1y}(v-v_0)\right]-\frac{2\pi k}{P}, \frac{2\pi}{\lambda}\left[d_{2x}(u-u_0)+d_{2y}(v-v_0)\right]-\frac{2\pi l}{Q}\right)\right|^2. \quad (22)$$

By analyzing (22), one can infer that

- the mainlobe of a *DS*-based array turns out to be in correspondence with the peak of (21) at $k = l = 0$, that is at the angular coordinates $(u,v) = (u_0, v_0)$ [Fig. 1(*b*)], and its value is
$$\mathcal{P}(u_0,v_0) = \mathcal{P}_{el}(u_0,v_0)\left[\gamma(PQ-1)+H\right]; \quad (23)$$
- the other (i.e., $k \times l \neq 0$) samples of $\mathcal{P}(u,v)$ are equal to (15)
$$\mathcal{P}(u_{kl},v_{kl}) = \mathcal{P}_{el}(u_{kl},v_{kl})(H-\gamma); \quad (24)$$
- the grating lobes (*GLs*) of *DS* thinned arrays appear at the angular coordinates [Fig. 8(*a*)]
$$(u_{bc}, v_{bc}) = \left\{u_0 + \lambda\frac{d_{2y}b - d_{1y}c}{\nu}, v_0 + \lambda\frac{d_{1x}c - d_{2x}b}{\nu}\right\} \quad b,\ c \in \mathbb{N},\ b \times c \neq 0 \quad (25)$$
since in (22) there are no other peaks of $\xi_{kl}$ except for $k = l = 0$, $|\exp(j\eta_{00})|^2 = 1$, and the periodicity of the grating lobes is that of $\mathcal{S}_{00}$, which is a periodic function in the $(\chi, \psi)$ domain (12) with period $2\pi$. These locations in the $(u,\ v)$-domain (25) are coincident with those of fully-populated layouts displaced over the same lattice [1] [Fig. 8(*b*)], thus *DS*-based arrays have the same field-of-view of standard arrangements.

Taking into account these considerations, the relation (22) can be used to derive closed-form expressions for the key figures of merit in synthesizing isophoric thinned arrays as detailed in the following sub-sections.

### A. Pattern Features for DS-Based Arrangements

*1) Sidelobe Level:* Starting from the definition of the sidelobe level

$$SLL \triangleq \frac{\max_{(u,v)\notin\mathcal{M}}\mathcal{P}(u,v)}{\mathcal{P}(u_0,v_0)} \quad (26)$$

$\mathcal{M}$ being the mainlobe region, only the numerator of (26) must be estimated since $\mathcal{P}(u_0,v_0)$ is already known (23). Towards this end, let us observe that a lower bound for $[\max_{(u,v)\notin\mathcal{M}}\mathcal{P}(u,v)]$ can be easily deduced by just observing that the sidelobes are necessarily above the values of $\mathcal{P}(u,v)$ at the known sampling points (24). Accordingly, one can write that

$$SLL > \frac{(H-\gamma)\epsilon}{\left[\gamma(PQ-1)+H\right]} \triangleq SLL_{INF}, \quad (27)$$

$\epsilon$ being a scaling term depending on the embedded element factor

$$\epsilon \triangleq \frac{\max_{k,\,l\,(k\neq l=0)}\left\{\mathcal{P}_{el}(u_{kl},v_{kl})\right\}}{\mathcal{P}_{el}(u_0,v_0)}, \quad (28)$$

by substituting (23) and (24) in (26).

On the other hand, the $SLL$ can be approximated - analogously as for standard/square-grid *DS* and *ADS* arrangements [16][17] - by setting the power pattern $\mathcal{P}(u,v)$ in (26) at the mid locations $(u_{mn},\ v_{mn})$

$$u_{mn} = u_0 + \lambda\frac{Q\left(m+\frac{1}{2}\right)d_{2y} - P\left(n+\frac{1}{2}\right)d_{1y}}{\nu PQ}$$
$$v_{mn} = v_0 + \lambda\frac{P\left(n+\frac{1}{2}\right)d_{1x} - Q\left(m+\frac{1}{2}\right)d_{2x}}{\nu PQ} \quad (29)$$

$(m = 0, ..., P-1;\ n = 0, ..., Q-1)$ between two adjacent known samples at $(u_{mn}, v_{mn})$ and $(u_{(m+1)(n+1)}, v_{(m+1)(n+1)})$: $SLL \approx \frac{\max_{(m,n)\notin\widehat{\mathcal{M}}}\mathcal{P}(u_{mn},v_{mn})}{\mathcal{P}(u_0,v_0)}$. Accordingly, it turns out that

$$SLL \approx \frac{1}{\mathcal{P}_{el}(u_0,v_0)\left[\gamma\times(P\times Q-1)+H\right]} \max_{(m,n)\notin\widehat{\mathcal{M}}}\left\{\mathcal{P}_{el}(u_{mn},v_{mn}) \times \left|\sqrt{\gamma(PQ-1)} \times \mathcal{S}\left(\frac{2\pi[d_{1x}(u_{mn}-u_0)+d_{1y}(v_{mn}-v_0)]}{\lambda}, \frac{2\pi[d_{2x}(u_{mn}-u_0)+d_{2y}(v_{mn}-v_0)]}{\lambda}\right) + \sqrt{H-\gamma}\sum_{k=0,\,l\neq 0}^{P-1}\sum_{l=0,\,k\neq 0}^{Q-1}\mathcal{S}\left(\frac{2\pi[d_{1x}(u_{mn}-u_0)+d_{1y}(v_{mn}-v_0)]}{\lambda}-\frac{2\pi k}{P}, \frac{2\pi[d_{2x}(u_{mn}-u_0)+d_{2y}(v_{mn}-v_0)]}{\lambda}-\frac{2\pi l}{Q}\right)\exp\left[j\eta_{kl}(u_{mn},v_{mn})\right]\right|^2\right\} \quad (30)$$

where $\widehat{\mathcal{M}}$ stands for the discrete version of $\mathcal{M}$ (see the Appendix A.3).

By substituting (29) in (12), it turns out that

$$\mathcal{S}\left(\frac{2\pi[d_{1x}(u_{mn}-u_0)+d_{1y}(v_{mn}-v_0)]}{\lambda}-\frac{2\pi k}{P}, \frac{2\pi[d_{2x}(u_{mn}-u_0)+d_{2y}(v_{mn}-v_0)]}{\lambda}-\frac{2\pi l}{Q}\right) = \frac{\sin\left[\pi\left(m+\frac{1}{2}-k\right)\right]}{P\sin\left[\frac{\pi}{P}\left(m+\frac{1}{2}-k\right)\right]}\frac{\sin\left[\pi\left(n+\frac{1}{2}-l\right)\right]}{Q\sin\left[\frac{\pi}{Q}\left(n+\frac{1}{2}-l\right)\right]} \times \exp\left[j\left(\pi\left(m+\frac{1}{2}-k\right)\frac{P-1}{P}+\pi\left(n+\frac{1}{2}-l\right)\frac{Q-1}{Q}\right)\right] \quad (31)$$

$[k = 0, ..., P-1;\ l = 0, ..., Q-1,\ (m,n) \notin \widehat{\mathcal{M}}]$ and after simple manipulations, since $\sin\left[\pi\left(m+\frac{1}{2}-k\right)\right] = (-1)^{m-k}$, $\sin\left[\pi\left(n+\frac{1}{2}-l\right)\right] = (-1)^{n-l}$, and also using the approximation

$$\exp\left[j\left(\frac{\pi(P-1)\left(m+\frac{1}{2}-k\right)}{P}+\frac{\pi(Q-1)\left(n+\frac{1}{2}-l\right)}{Q}\right)\right] \approx \approx (-1)^{m+n-k-l} \quad (32)$$

valid for large $P$ and/or $Q$ values [17], one can obtain the following expression

$$\mathcal{S}\left(\frac{2\pi[d_{1x}(u_{mn}-u_0)+d_{1y}(v_{mn}-v_0)]}{\lambda}-\frac{2\pi k}{P}, \frac{2\pi[d_{2x}(u_{mn}-u_0)+d_{2y}(v_{mn}-v_0)]}{\lambda}-\frac{2\pi l}{Q}\right) = \left\{PQ\sin\left[\frac{\pi}{P}\left(m-k+\frac{1}{2}\right)\right]\sin\left[\frac{\pi}{Q}\left(n-l+\frac{1}{2}\right)\right]\right\}^{-1}. \quad (33)$$

Next, (33) is inserted in (30) and then

$$SLL \approx \frac{1}{\mathcal{P}_{el}(u_0,v_0)\times[\gamma(PQ-1)+H]} \max_{(m,n)\notin\widehat{\mathcal{M}}}\left\{\mathcal{P}_{el}(u_{mn},v_{mn})\times \left|\frac{\sqrt{\gamma(PQ-1)}}{P\times Q\times\sin\left[\frac{\pi}{P}\left(m+\frac{1}{2}\right)\right]\sin\left[\frac{\pi}{Q}\left(n+\frac{1}{2}\right)\right]}+ +\sqrt{H-\gamma}\sum_{k=0,\,l\neq 0}^{P-1}\sum_{l=0,\,k\neq 0}^{Q-1}\frac{\exp[j\eta_{kl}]}{PQ\sin\left[\frac{\pi}{P}\left(m-k+\frac{1}{2}\right)\right]\sin\left[\frac{\pi}{Q}\left(n-l+\frac{1}{2}\right)\right]}\right|^2\right\}. \quad (34)$$

Since the first term in (34) is negligible outside the mainlobe region $\mathcal{M}$ (Appendix A.3), (34) becomes

$$SLL\{\mathcal{F}(u,v)\} \approx \frac{H-\gamma}{\mathcal{P}_{el}(u_0,v_0)PQ[\gamma(PQ-1)+H]}\times \max_{(m,n)\notin\widehat{\mathcal{M}}}\left\{\mathcal{P}_{el}(u_{mn},v_{mn})\times \times\left|\sum_{k=0,\,l\neq 0}^{P-1}\sum_{l=0,\,k\neq 0}^{Q-1}\frac{\exp[j\eta_{kl}]}{\sin\left[\frac{\pi}{P}\left(m-k+\frac{1}{2}\right)\right]\sin\left[\frac{\pi}{Q}\left(n-l+\frac{1}{2}\right)\right]}\right|^2\right\} \quad (35)$$

and it can be upper bounded by using the Monte Carlo procedure adopted in [16][17] to yield

$$\left|\frac{1}{PQ}\sum_{k=0,\,l\neq 0}^{P-1}\sum_{l=0,\,k\neq 0}^{Q-1}\left[\frac{\exp[j\eta_{kl}]}{\sin\left[\frac{\pi}{P}\left(m-k+\frac{1}{2}\right)\right]\sin\left[\frac{\pi}{Q}\left(n-l+\frac{1}{2}\right)\right]}\right]\right|^2 < < 0.5+1.5\log_{10}(PQ) \quad (36)$$

that holds true for at least one binary sequence $\underline{\alpha}$ among the cyclic shifted versions $\underline{\alpha}^{(\sigma_x,\sigma_y)}$ of the reference *DS*, and finally

$$SLL < \frac{\epsilon(H-\gamma)[0.5+1.5\log_{10}(PQ)]}{\gamma(PQ-1)+H} \triangleq SLL_{SUP}. \quad (37)$$

By combining (27) and (37), the following closed-form bounds are derived

$$SLL_{INF} \leq SLL \leq SLL_{SUP}. \quad (38)$$

*2) Directivity:* The customization of the expression of the maximum directivity

$$D = \frac{4\pi\mathcal{P}(u_0,v_0)}{\int_\Phi\left[\mathcal{P}(u,v)\sin\left(\sqrt{u^2+v^2}\right)\frac{\partial(\theta,\varphi)}{\partial(u,v)}\right]dudv}, \quad (39)$$

where $\frac{\partial(\theta,\varphi)}{\partial(u,v)}$ and $\Phi$ are the Jacobian of the $(\theta,\varphi) \leftrightarrow (u,v)$ transformation and the visible range ($\Phi \triangleq \{(u,v): u^2+v^2\leq 1\}$), respectively, to *DS*-based thinned arrays needs the evaluation of the integral at the denominator of (39), but this is not possible without *a-priori* knowing $\mathcal{P}_{el}(u,v)$ (i.e., the single radiator of the array) and $\eta_{kl}$ (i.e., the sequence $\underline{\alpha}$), while the following lower bound can be deduced[4] (see the Appendix A.4)

$$\int_\Phi\left[\mathcal{P}(u,v)\sin\left(\sqrt{u^2+v^2}\right)\frac{\partial(\theta,\varphi)}{\partial(u,v)}\right]\mathrm{d}u\mathrm{d}v \leq \leq 2\pi\left\{P\times Q\times\gamma\left[1-\cos\left(\overline{\theta}\right)\right]+H-\gamma\right\} \quad (40)$$

where $\overline{\theta}$ is reported in the Appendix. By using (23) and exploiting the inequality in (40), it turns out that

$$D \geq D_{INF} \quad (41)$$

[4]An upper bound for $D$ cannot be easily deduced since the single element directivity is in principle not limited. On the other hand, it is worth pointing out that such an upper bound is not required for the *Isophoric Thinning Problem* as shown in the following (Sect. IV-B).

where

$$D_{INF} \triangleq \frac{2[\gamma(PQ-1)+H]}{\gamma\left\{PQ\left[1-\cos\left(\overline{\theta}\right)\right]-1\right\}+H}. \quad (42)$$

*3) Setup of a Pattern Value/Threshold along a User-Defined Angular Direction :* In order to set a user-defined pattern value or threshold, $\mathcal{P}^T$, along an angular direction of interest, $(u^T,v^T)$, it is enough to force the two following conditions:

- the power pattern $\mathcal{P}(u,v)$ along the direction, $(u^T,v^T)$, has a magnitude (24) equal or smaller than a user-defined value $\mathcal{P}^T$

$$(H-\gamma)\times\mathcal{P}_{el}\left(u^T,v^T\right) \leq \mathcal{P}^T; \quad (43)$$

- the angular direction $(u_{mn},v_{mn})$ (8), which is a function of the lattice descriptors, $(d_{1x},\,d_{1y},\,d_{2x},\,d_{2y})$, coincides with the target direction

$$\left(u_0+\frac{\lambda}{PQ}\frac{Qmd_{2y}-Pnd_{1y}}{\nu},\,v_0+\frac{\lambda}{PQ}\frac{Pnd_{1x}-Qmd_{2x}}{\nu}\right) = =\left(u^T,v^T\right) \quad (44)$$

$\{m,\,n\}$ being a couple of arbitrary user-chosen integers within the admissibility ranges $m\in[0,P-1]$ and $n\in[0,Q-1]$ subject to $m\times n\neq 0$;

*4) Maximum Half-Power Beamwidth:* The computation of the maximum half-power beamwidth, which is defined as

$$BW^{\max} = \max_{\overline{\varphi}\in[0,2\pi]}\left\{BW_{\overline{\varphi}}\left[\mathcal{P}(u,v)\right]\right\} \quad (45)$$

$BW_{\overline{\varphi}}[\mathcal{P}(u,v)]$ being the half-power beamwidth along the $\varphi=\overline{\varphi}$ angular cut, can be performed by exploiting the well-known relation between array directivity and beamwidth [1]

$$D \approx \frac{4\pi(0.886)^2}{(BW^{\max})^2} \quad (46)$$

and using the directivity inequality, $D\geq D_{INF}$, from Sect. IV-A2. The result is that

$$BW^{\max} \leq BW_{SUP} \quad (47)$$

where $BW_{SUP} = 0.886\sqrt{\frac{2\pi\left\{\gamma\times\left\{P\times Q\times\left[1-\cos\left(\overline{\theta}\right)\right]-1\right\}+H\right\}}{\gamma\times(P\times Q-1)+H}}$.

## B. Analytic-Method for Isophoric Thinning

Let us now detail the original formulation of the synthesis of two-dimensional isophoric thinned arrays (“*to find the aperture size* $P\times Q$*, the unit cell axes* $\mathbf{d}_1$ *and* $\mathbf{d}_2$*, and the binary sequence* $\underline{\alpha}$ *so that the figures of merit of the corresponding radiated pattern* $\mathcal{P}(u,v)$ *satisfy user-defined constraints*”) as follows “*to find the aperture size* $P\times Q$*, the unit cell axes* $\mathbf{d}_1$ *and* $\mathbf{d}_2$, and *a* $(P\times Q,H,\gamma)$*-DS so that the corresponding radiated pattern* $\mathcal{P}(u,v)$ *has:* (1) *sidelobe level smaller than* $SLL^T$, (2) *a directivity greater than* $D^T$, (3) *a pattern value along the angular direction* $(u^T,v^T)$ *equal or smaller than* $\mathcal{P}^T$*, and* (4) *a maximum half-power beamwidth smaller than* $BW^T$”. Taking into account the closed-form expressions for the key figures of merit of *DS*-based thinned arrays (38)(41)(44)(43)(47), the solution to this *Isophoric Thinning Problem* can be analytically found by identifying $\underline{\alpha}$ (i.e., admissible values of $P$, $Q$, $H$, and $\gamma$ from *DS* repositories [28][29]) and $\mathbf{d}_1$, $\mathbf{d}_2$ such that

$$\begin{cases} SLL_{SUP} \leq SLL^T \\ D_{INF} \geq D^T \\ \left(u_0 + \frac{\lambda}{PQ}\frac{mQd_{2y}-nPd_{1y}}{\nu},\, v_0 + \frac{\lambda}{PQ}\frac{nPd_{1x}-mQd_{2x}}{\nu}\right) = \left(u^T, v^T\right) \\ (H-\gamma) \times \mathcal{P}_{el}\left(u^T, v^T\right) \leq \mathcal{P}^T \\ \sqrt{u_{bc}^2 + v_{bc}^2} > 1 \\ BW_{SUP} \leq BW^T. \end{cases} \tag{48}$$

More in detail, the following design procedure can be derived from (48):

- ***Step 1***: *Identification of the* $(P \times Q, H, \gamma)$*-DS Generator*
  Identify the admissible values of $P$, $Q$, $H$, $\gamma$ either from *DS* repositories [28][29] or constructed through ad-hoc methodologies [30] such that the condition $SLL_{SUP} \leq SLL^T$ (38) is satisfied;
- ***Step 2***: *Identification of the Lattice*
  If $\mathcal{P}_{el}\left(u^T, v^T\right) \leq \frac{\mathcal{P}^T}{(H-\gamma)}$ (43) then set $\mathbf{d}_1$ and $\mathbf{d}_2$ so that $\left(u_0 + \frac{\lambda}{PQ}\frac{mQd_{2y}-nPd_{1y}}{\nu},\, v_0 + \frac{\lambda}{PQ}\frac{nPd_{1x}-mQd_{2x}}{\nu}\right) = \left(u^T, v^T\right)$ (44) subject to the *GL*s-free condition ($\sqrt{u_{bc}^2 + v_{bc}^2} > 1$, $\forall\, b, c \in \{-1, 0, 1\}$) else change the radiating element [i.e., $\mathcal{P}_{el}(u, v)$] or repeat ***Step 1***;
- ***Step 3***: *Check Bounds Compliance*
  Substitute the values of $P$, $Q$, $H$, $\gamma$, $\mathbf{d}_1$, and $\mathbf{d}_2$ selected on ***Step 1*** and ***Step 2*** in (41) and (47), if $D_{INF} \geq D^T$ and $BW_{SUP} \leq BW^T$ goto ***Step 4***, else repeat ***Step 2*** until ($D_{INF} \geq D^T$ and $BW_{SUP} \leq BW^T$) otherwise repeat ***Step 1***;
- ***Step 4***: *Identification of the DS Sequence*
  Cyclically-shift the reference *DS* layout to find the best solution, $\underline{\alpha}^{(\sigma^{opt})}$, fitting the user-defined constraints and minimizing the *SLL*: $\sigma^{opt} = \arg\left[\min_\sigma\left(SLL^{(\sigma)}\right)\right]$ being $\sigma \triangleq \sigma_x Q + \sigma_y$;

As for the closed-form design equations in (48) and the arising synthesis procedure, it is worthwhile to remark that (*i*) the pattern control is yielded without considering tapering/sparsening, thus minimizing the realization complexity of the array layout; (*ii*) according to the derivation in Sect. IV-A1, a *DS* layout that complies with the *a-priori SLL* bound can be determined by simulating and comparing the $P \times Q$ cyclic shifted versions of the reference *DS* sequence $\underline{\alpha}^{(\sigma)}$; (*iii*) except for the *SLL*, all the cyclic shifts of the *DS* generator sequence, which fits the design-constraints in (48), are still suitable solutions for the problem at hand. This outcome underlines the fact that designing an isophoric thinned array is equivalent to set a suitable autocorrelation function (19) whose entries are univocally determined by the *DS* descriptors (i.e., $P \times Q$, $H$, $\gamma$); (*iv*) the analytic relations derived in [16][17] are the customization to square lattices of the more general theory presented in this paper.

## V. Numerical Assessment

The objective of this section is twofold. On the one hand, it is aimed at assessing the reliability of (*i*) the closed-form expressions derived in Sect. II to predict the power pattern $\mathcal{P}(u, v)$ generated by any excitation sequence $\underline{\alpha}$ and of (*ii*) the analytic bounds (38), (41), (44), and (47) for the key figures of merit of thinned *DS*-based arrays displaced over arbitrary lattices. On the other hand, it is devoted to give some insights on the potentialities and the features of the analytical design strategy summarized in (48). Towards this end, different numerical examples dealing with various apertures, lattice geometries, and values of the thinning factor $\tau$ ($\tau \triangleq \frac{H}{PQ}$) will be considered in the following.

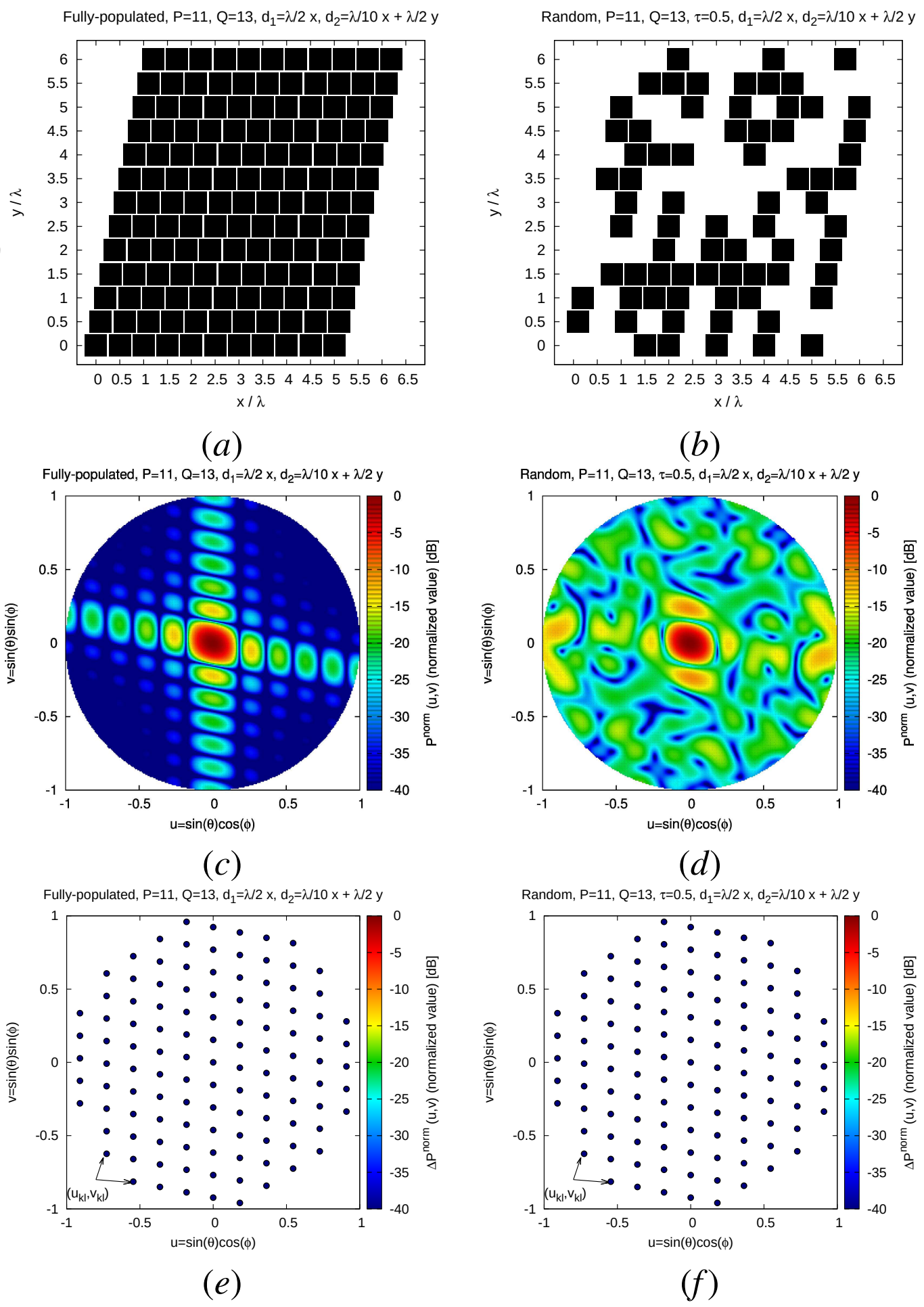

Figure 2. *Pattern Prediction* ($P = 11$, $Q = 13$, $\mathbf{d}_1 = \frac{\lambda}{2}\widehat{\mathbf{x}}$, $\mathbf{d}_2 = \left(\frac{\lambda}{10}\widehat{\mathbf{x}} + \frac{\lambda}{2}\widehat{\mathbf{y}}\right)$, $\mathcal{E}(u, v) = 1.0$, $\theta_0 = \varphi_0 = 0.0$ [deg]) - Plots of (*a*)(*b*) the layout, (*c*)(*d*) the normalized power pattern, $\mathcal{P}^{norm}(u, v)$, and (*e*)(*f*) predicted pattern sample differences, $\Delta\mathcal{P}^{norm}(u_{kl}, v_{kl})$, $k = 0, ..., (P-1)$; $l = 0, ..., (Q-1)$ (15) of (*a*)(*c*)(*e*) the fully-populated array and (*b*)(*d*)(*f*) the randomly thinned ($\tau = 0.5$) one.

### A. Pattern Prediction for Arbitrary Excitations Sequences $\underline{\alpha}$

The purpose of the first numerical experiment is the validation of the generality of the expression (15) in predicting the samples of $\mathcal{P}(u, v)$ along the directions $(u_{kl}, v_{kl})$, $k = 0, ..., (P-1)$; $l = 0, ..., (Q-1)$ (8) for any excitation sequence $\underline{\alpha}$ and lattice unit cell. Towards this end, an aperture of $P \times Q = 11 \times 13$ isotropic [$\mathcal{E}(u, v) = 1.0$] elements steered along $\theta_0 = \varphi_0 = 0.0$ [deg] is considered as a benchmark, while a fully-populated layout [Fig. 2(*a*)] and a randomly thinned arrangement [Fig. 2(*b*)] lying on a non-rectangular grid [$\mathbf{d}_1 = \frac{\lambda}{2}\widehat{\mathbf{x}}$ and $\mathbf{d}_2 = \left(\frac{\lambda}{10}\widehat{\mathbf{x}} + \frac{\lambda}{2}\widehat{\mathbf{y}}\right)$] have been taken into account as representative of a general class of arrays. Figures 2($c$)-2($d$) show the plots of the normalized power patterns, $\mathcal{P}^{norm}(u,v)$ [$\mathcal{P}^{norm}(u,v) \triangleq \frac{\mathcal{P}(u,v)}{\mathcal{P}(u_0,v_0)}$], while Figs. 2($e$)-2($f$) give the values of their deviation with respect to the *DFT* of the cyclic autocorrelation $\underline{a}$ of the corresponding weighting sequences $\underline{\alpha}$ (i.e., $\Delta\mathcal{P}^{norm}(u_{kl},v_{kl}) \triangleq \frac{\mathcal{P}(u_{kl},v_{kl})-\xi_{kl}}{\mathcal{P}(u_0,v_0)}$). As it can be observed and theoretically expected, the values of $\mathcal{P}^{norm}(u,v)$ at the angular coordinates $(u_{kl},v_{kl})$ (8) [Figs. 2($c$)-2($d$)] coincide with the predictions through (15) [Figs. 2($e$)-2($f$)] even though non-rectangular unit cells are at hand. Since this property is not related to the features/properties of the excitation sequence under analysis [e.g., uniform - Fig. 2($c$) vs. Fig. 2($e$) - or random - Fig. 2($d$) vs. Fig. 2($f$) - excitations], it opens the doors to a new theoretical tool for the design of array layouts whenever ($i$) either $\underline{a}$ is *a-priori* known/can be estimated (e.g., $\alpha$ is a sequence with known autocorrelation functions) or ($b$) $\underline{a}$ has to be determined as another degree-of-freedom for fitting user-defined pattern requirements/constraints.

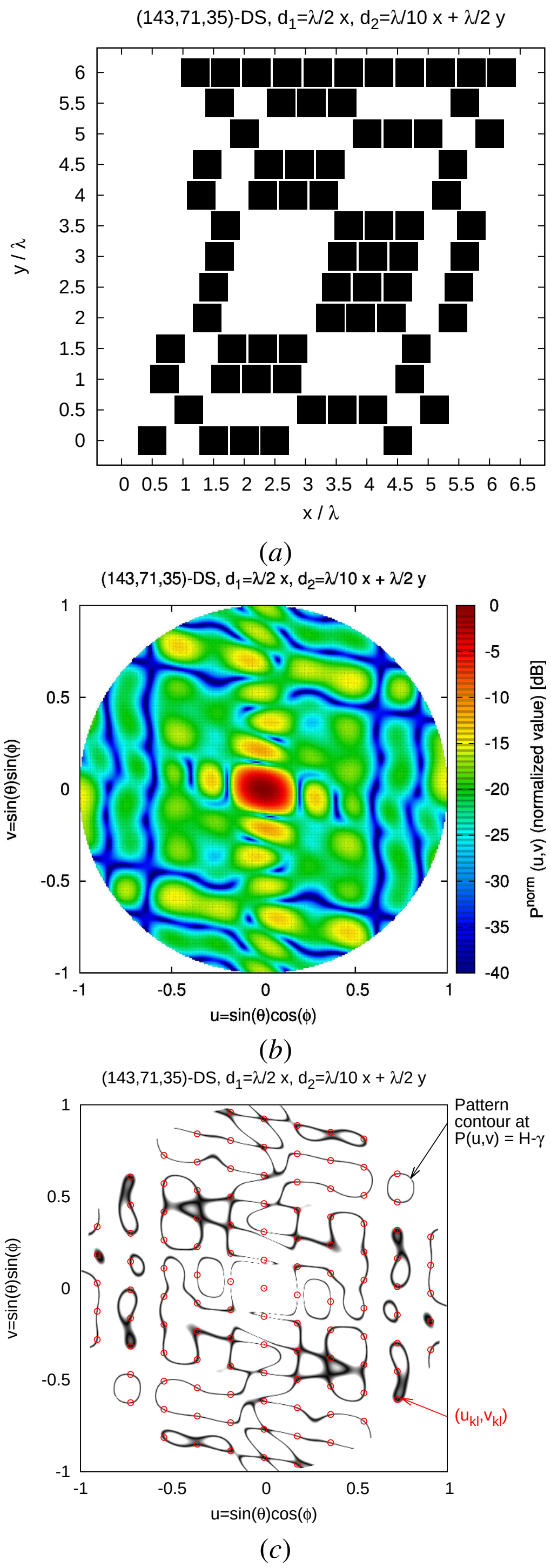

Figure 3. *Performance Bounds* ($P = 11$, $Q = 13$, $H = 71$, $\gamma = 35$, $\mathbf{d}_1 = \frac{\lambda}{2}\widehat{\mathbf{x}}$, $\mathbf{d}_2 = \left(\frac{\lambda}{10}\widehat{\mathbf{x}} + \frac{\lambda}{2}\widehat{\mathbf{y}}\right)$, $\mathcal{E}(u,v) = 1.0$, $\theta_0 = \varphi_0 = 0.0$ [deg]) - Plots of ($a$) the array layout, ($b$) the normalized power pattern, $\mathcal{P}^{norm}(u,v)$, and ($c$) the pattern contour, $\mathcal{P}^{norm}_{kl}$, together with the predicted power samples, $\mathcal{P}^{norm}(u_{kl},v_{kl})$, $k = 0,...,(P-1)$; $l = 0,...,(Q-1)$ (23)(24).

Table I
*DS Sequences* - DESCRIPTIVE PARAMETERS.

| $P$ | $Q$ | $H$ | $\gamma$ | $\tau$ | Ref. |
|---|---|---|---|---|---|
| 8 | 8 | 36 | 20 | 0.562 | [29] |
| 11 | 13 | 71 | 35 | 0.496 | [28] |
| 12 | 12 | 78 | 42 | 0.541 | [29] |
| 15 | 17 | 127 | 63 | 0.498 | [28] |
| 16 | 16 | 136 | 72 | 0.531 | [29] |
| 17 | 19 | 161 | 80 | 0.498 | [28] |
| 31 | 33 | 511 | 255 | 0.499 | [28] |

## *B. Pattern Prediction and Validation of* a-priori *Performance Bounds of* DS*-based Arrangements*

Dealing with *DS*-arrays, the expressions (14) and (15) customize into (22) and (23)(24). The next experiment is then aimed at assessing the reliability of these latter also pointing out the possibility to *a-priori* predict the pattern samples of isophoric arrangements based on *DS* excitations and displaced over arbitrary lattices from just the knowledge of the *DS* descriptors (i.e., $P$, $Q$, $H$, and $\gamma$). Towards this end, a $P \times Q = 11 \times 13$ layout generated from a $(143, 71, 35)$-*DS* [28] (i.e., $H = 71$, $\gamma = 35$ - Tab. I) with axes of the array lattice equal to [$\mathbf{d}_1 = \frac{\lambda}{2}\widehat{\mathbf{x}}$, $\mathbf{d}_2 = \left(\frac{\lambda}{10}\widehat{\mathbf{x}} + \frac{\lambda}{2}\widehat{\mathbf{y}}\right)$] has been considered [Fig. 3($a$)]. According to (22), the pattern samples at (8) ($u_{kl} = 1.82 \times 10^{-1}k$, $v_{kl} = -3.63 \times 10^{-2}k + 1.53 \times 10^{-1}l$) are expected to assume the values (23) $\mathcal{P}(u_{kl},v_{kl}) = \gamma \times (P \times Q - 1) + H = 5041$ when $k = l = 0$, and (24) $\mathcal{P}(u_{kl},v_{kl}) = H - \gamma = 36$ for $k \times l \neq 0$. To validate such predictions, let us check whether the normalized pattern $\mathcal{P}^{norm}(u,v)$ [Fig. 3($b$)] crosses the normalized pattern *contour* $\mathcal{P}^{norm}_{kl}$ ($\mathcal{P}^{norm}_{kl} \triangleq \frac{\mathcal{P}(u_{kl},v_{kl})}{\mathcal{P}(u_0,v_0)}$, $\mathcal{P}^{norm}_{kl} = \frac{H-\gamma}{\gamma\times(P\times Q-1)+H} = -21.46$ [dB]) at the angular coordinates equal to $(u_{kl},v_{kl})$. This is visually confirmed in Fig. 3($c$) where the red dots, which identify the intersections of the black contour of $\mathcal{P}^{norm}_{kl}$ with the plot of $\mathcal{P}^{norm}(u,v)$, exactly lie at $(u,v) = (u_{kl},v_{kl})$. Theoretically, it should work changing the lattice by enlarging and/or rotating the lattice grid, as well. This can be easily proved by referring to the same *DS* arrangement, but varying the values of $\mathbf{d}_1$ and $\mathbf{d}_2$ of the lattice cell. For instance, let us consider the array layouts yielded in correspondence with the same *DS* sequence, while setting $\mathbf{d}_1=\frac{\lambda}{2}\widehat{\mathbf{x}}$, $\mathbf{d}_2=\left(\frac{3\lambda}{10}\widehat{\mathbf{x}}+\frac{\lambda}{2}\widehat{\mathbf{y}}\right)$ [Fig. 4$(a)$] and $\mathbf{d}_1=\frac{\lambda}{2}\widehat{\mathbf{x}}$, $\mathbf{d}_2=\left(\frac{\lambda}{2}\widehat{\mathbf{x}}+\frac{\lambda}{2}\widehat{\mathbf{y}}\right)$ [Fig. 4$(b)$]. Of course, the corresponding power patterns [Figs. 4$(c)$-4$(d)$] as well as the angular values $(u_{kl},v_{kl})$ significantly change from that one in Fig. 3$(b)$ due to the lattice variations. More specifically, it turns out that $u_{kl}=1.82\times10^{-1}k$ and $v_{kl}=-1.09\times10^{-1}k+1.53\times10^{-1}l$ [Fig. 4$(e)$] and $u_{kl}=1.82\times10^{-1}k$, $v_{kl}=-1.82\times10^{-1}k+1.53\times10^{-1}l$ [Fig. 4$(f)$] according to the dependence of the angular coordinates on the lattice descriptors (8). Nevertheless, the pattern samples, $\mathcal{P}(u_{kl},v_{kl})$, whose magnitudes do not depend on $\mathbf{d}_1$, $\mathbf{d}_2$ (23)(24), can be still reliably predicted as shown in Figs. 4$(e)$-4$(f)$ where the same pictorial representation/meaning of Fig. 3$(c)$ has been used. Starting from these outcomes, it is then worth noticing that the unit cell has a direct impact on the position of the predicted samples, but neither on their values nor on their predictability, this implies that a desired pattern value can be enforced in any desired direction, $(u^T,v^T)$, by simply modifying the lattice of the *DS* arrangement through (44).

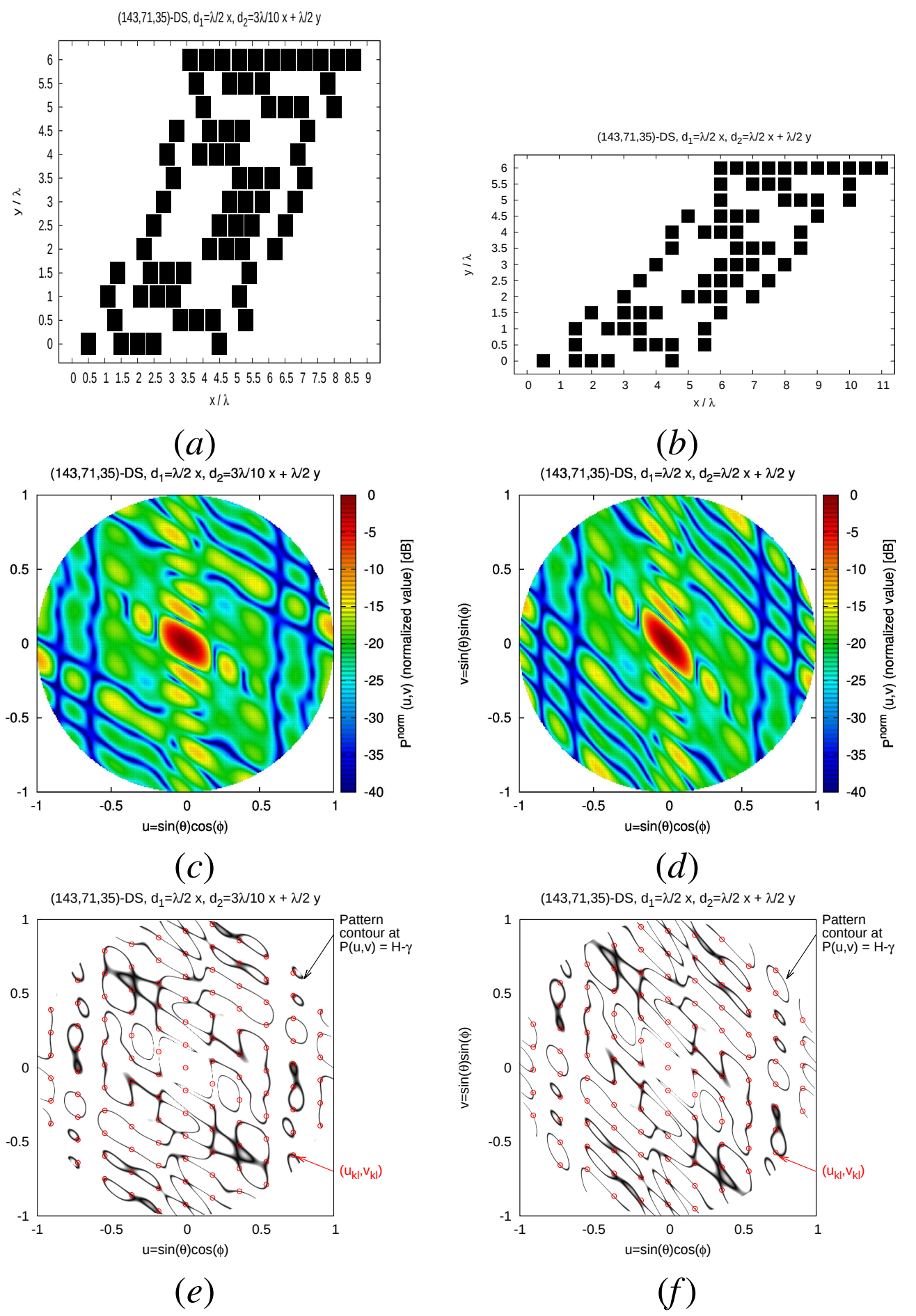

Figure 4. *Performance Bounds* ($P=11$, $Q=13$, $H=71$, $\gamma=35$, $\mathcal{E}(u,v)=1.0$, $\theta_0=\varphi_0=0.0$ [deg]) - Plots of $(a)(b)$ the array layout, $(c)(d)$ the normalized power pattern, $\mathcal{P}^{norm}(u,v)$, and $(e)(f)$ the pattern contour, $\mathcal{P}^{norm}_{kl}$, together with the predicted power samples, $\mathcal{P}^{norm}(u_{kl},v_{kl})$, $k=0,...,(P-1)$; $l=0,...,(Q-1)$ (23)(24) when $(a)(c)(e)$ $\mathbf{d}_1=\frac{\lambda}{2}\widehat{\mathbf{x}}$, $\mathbf{d}_2=\left(\frac{3\lambda}{10}\widehat{\mathbf{x}}+\frac{\lambda}{2}\widehat{\mathbf{y}}\right)$ and $(b)(d)(f)$ $\mathbf{d}_1=\frac{\lambda}{2}\widehat{\mathbf{x}}$, $\mathbf{d}_2=\left(\frac{\lambda}{2}\widehat{\mathbf{x}}+\frac{\lambda}{2}\widehat{\mathbf{y}}\right)$.

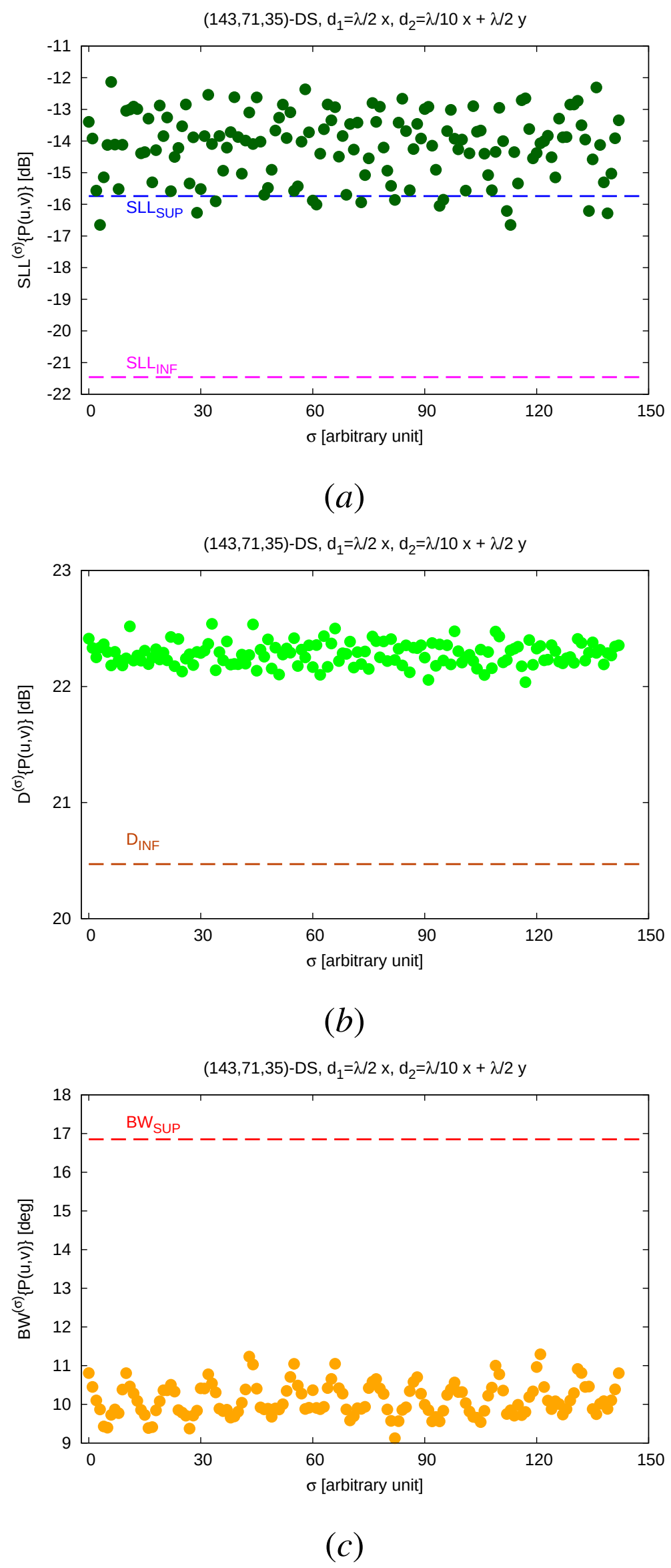

Figure 5. *Performance Bounds* ($P=11$, $Q=13$, $H=71$, $\gamma=35$, $\mathbf{d}_1=\frac{\lambda}{2}\widehat{\mathbf{x}}$, $\mathbf{d}_2=\left(\frac{\lambda}{10}\widehat{\mathbf{x}}+\frac{\lambda}{2}\widehat{\mathbf{y}}\right)$, $\mathcal{E}(u,v)=1.0$, $\theta_0=\varphi_0=0.0$ [deg]) - Behaviour of $(a)$ the $SLL$ $(b)$ the $D$, and $(c)$ the $BW^{\max}$ of isophoric arrays derived from the *DS* sequence $\underline{\alpha}^{(\sigma)}$ versus the shift index $\sigma$.

While (23), (24), and (8) state the possibility to predict the pattern samples of *DS*-based arrangements, they have been also used in Sect. IV to derive analytic bounds for the corresponding key pattern indexes. For assessment purposes, the values of the *SLL* [Fig. 5$(a)$], the directivity $D$ [Fig. 5$(b)$], and the maximum half-power beamwidth $BW^{max}$ [Fig. 5$(c)$] of the *DS*-based arrays derived from the reference layout in Fig. 3$(a)$ and its $(\sigma_x,\sigma_y)$ cyclic shifts are reported in Fig. 4 versus the shift index $\sigma$. By comparing the behavior of $SLL^{(\sigma)}$, $D^{(\sigma)}$, and $BW^{(\sigma)}$ with the *a-priori* bounds equal to $-21.46\,[\text{dB}]\leq SLL\leq-15.74\,[\text{dB}]$ (38), $D>20.47$ [dB] (41), and $BW^{\max}\leq16.85$ [deg] (47), respectively, it turns out that: $(a)$ as theoretically expected ("*at least one binary sequence* $\underline{\alpha}$*, among the cyclic shifted versions* $\underline{\alpha}^{(\sigma)}$ *of the reference DS, fits* (38)") different trade-off layouts (i.e., several $\sigma$ values) verify (38) [Fig. 5($a$)]; ($b$) all shifted sequences comply with (41) being $D^{(\sigma)}\geq D_{INF}$ [Fig. 5($b$)]; ($c$) the lower bound $D_{INF}$ is more than 1.0 dB below the actual $D^{(\sigma)}$ values because of the underestimation procedure employed to derive (41). This implies that the upper bound for the half-power beamwidth, $BW_{SUP}$, is overestimated and (47) widely holds true for all $\sigma$-shifts of the *DS* generator sequence [Fig. 5($c$)]; ($d$) the bounds deduced in Sect. IV-A for *DS* isophoric thinned arrays hold true even though the aperture at hand is not very large [e.g., $P\times Q=11\times 13$ - Fig. 3($a$)].

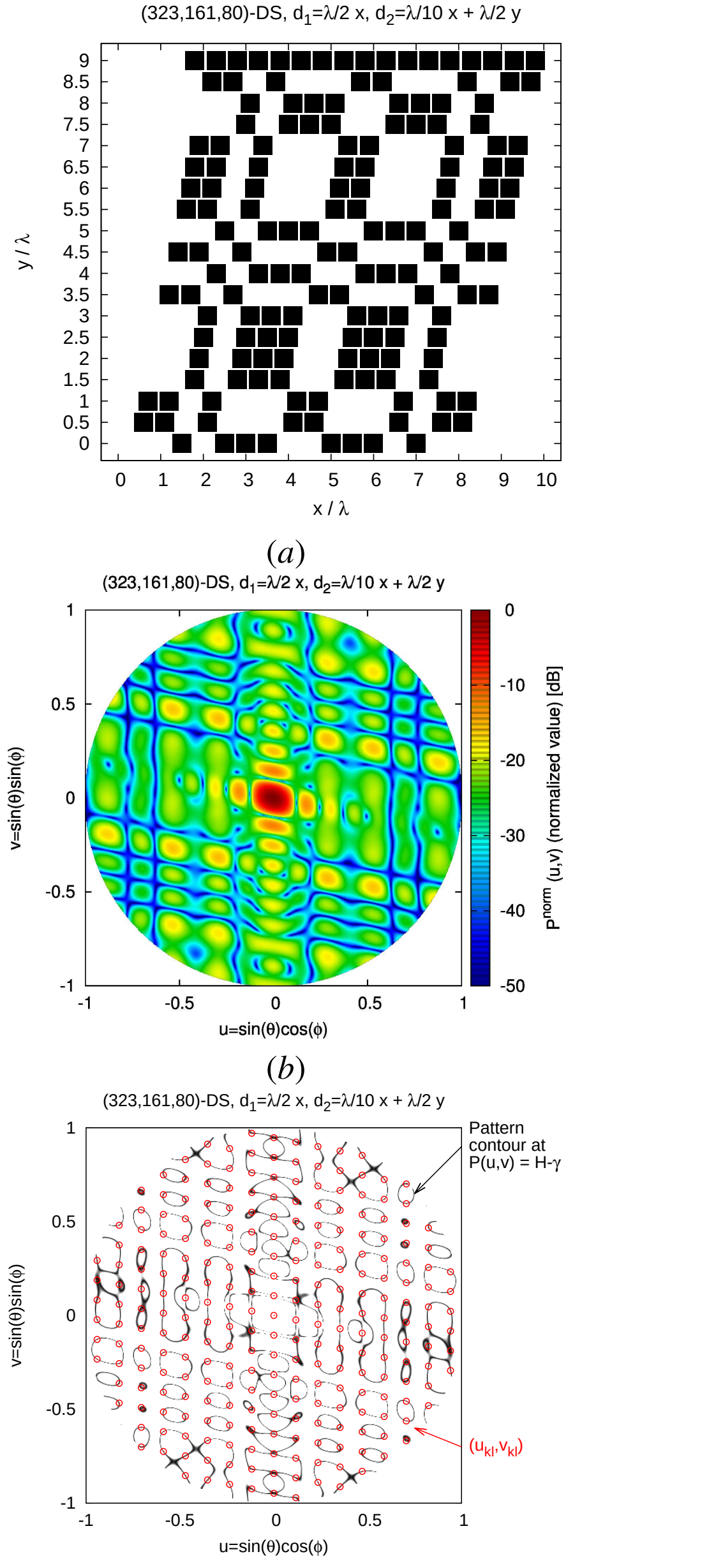

Figure 6. *Performance Bounds* ($P=17$, $Q=19$, $H=161$, $\gamma=80$, $\mathbf{d}_1=\frac{\lambda}{2}\widehat{\mathbf{x}}$, $\mathbf{d}_2=\left(\frac{\lambda}{10}\widehat{\mathbf{x}}+\frac{\lambda}{2}\widehat{\mathbf{y}}\right)$, $\mathcal{E}(u,v)=1.0$, $\theta_0=\varphi_0=0.0$ [deg]) - Plots of ($a$) the array layout, ($b$) the normalized power pattern, $\mathcal{P}^{norm}(u,v)$, and ($c$) the pattern contour, $\mathcal{P}^{norm}_{kl}$, together with the predicted power samples, $\mathcal{P}^{norm}(u_{kl},v_{kl})$, $k=0,...,(P-1)$; $l=0,...,(Q-1)$ (23)(24).

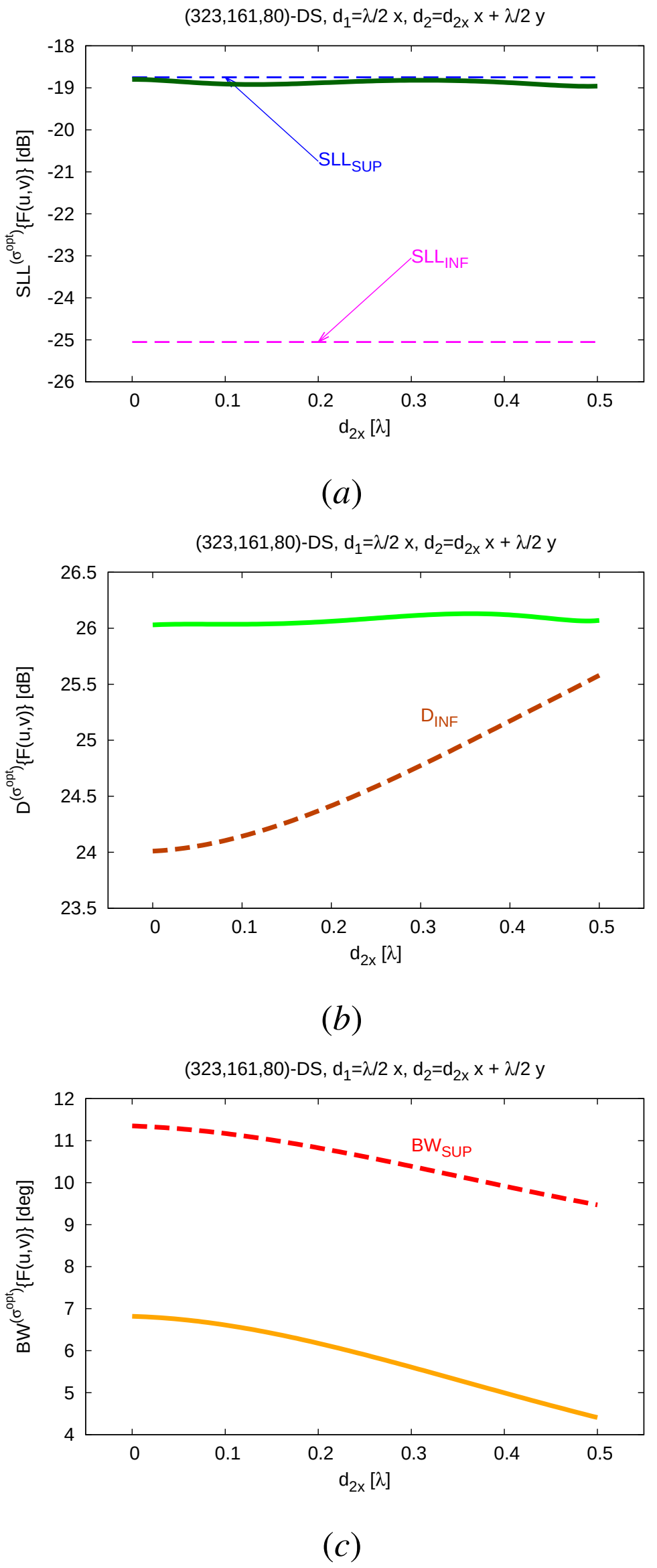

Figure 7. *Performance Bounds* ($P=17$, $Q=19$, $H=161$, $\gamma=80$, $\mathbf{d}_1=\frac{\lambda}{2}\widehat{\mathbf{x}}$, $\mathbf{d}_2=\left(d_{2x}\widehat{\mathbf{x}}+\frac{\lambda}{2}\widehat{\mathbf{y}}\right)$, $\mathcal{E}(u,v)=1.0$, $\theta_0=\varphi_0=0.0$ [deg]) - Behaviour of ($a$) $SLL^{(\sigma^{opt})}$ ($b$) $D^{(\sigma^{opt})}$, and ($c$) $BW^{(\sigma^{opt})}$ versus the value of the lattice side $d_{2x}$.

By keeping the same unit cell [i.e., $\mathbf{d}_1=\frac{\lambda}{2}\widehat{\mathbf{x}}$, $\mathbf{d}_2=\left(\frac{\lambda}{10}\widehat{\mathbf{x}}+\frac{\lambda}{2}\widehat{\mathbf{y}}\right)$], the case of a wider aperture is then analyzed. With reference to the $(323,161,80)$-*DS* [28] (i.e., $P=17$, $Q=19$, $H=161$, $\gamma=80$, and $\tau\approx 0.5$ - Tab. I), the power pattern [Fig. 6($b$)] radiated by the corresponding isophoric layout [Fig. 6($a$)] shows a regular sidelobe behaviour as predicted by (24). Moreover, the plot of the normalized pattern *contour* in Fig. 6($c$) confirms that the pattern samples with value $\mathcal{P}^{norm}_{kl}$ ($\mathcal{P}^{norm}_{kl}=\frac{H-\gamma}{\gamma\times(P\times Q-1)+H}=-25.05$ [dB]) are correctly foreseen, according to (24), along the cosine directions (8) ($u_{kl}=1.17\times 10^{-1}k$, $v_{kl}=-2.35\times 10^{-2}k+1.05\times 10^{-1}l$). As for the prediction of $SLL$, $D$, and $BW^{\max}$, Figure 7 compares the behaviour of the sidelobe level [Fig. 7($a$)], the directivity [Fig. 7($b$)], and the maximum half-power beamwidth [Fig. 7($c$)] of the optimal (i.e., $\sigma=\sigma^{opt}$, $\sigma^{opt} = \arg\left[\min_{\sigma}\left(SLL^{(\sigma)}\right)\right]$) *DS*-shifted arrangement with the closed-form bounds (38)(41)(47) when varying the lattice shape by changing the grid side $d_{2x}$ in the range $d_{2x} \in \left[0, \frac{\lambda}{2}\right]$. Once more, (*i*) the sidelobe level complies with the bounds (38) for any considered lattice [Fig. 7(*a*)], (*ii*) the array directivity always agrees with the lower bound (41) that, as in the previous cases, underestimates the actual $D^{\left(\sigma^{opt}\right)}$ [Fig. 7(*b*)], and (*iii*) the half-power beamwidth is correctly upper bounded by (47) [Fig. 7(*c*)].

Another interesting result of the general theory for *DS*-based arrays developed in (IV) is the feature of *a-priori* determining, thanks to (25), the grating lobe positions of the arising layout as well as, indirectly, the steering range/field of view of the *DS*-based isophoric arrays. As a matter of fact, the plot of $\mathcal{P}^{norm}(u,v)$ in a domain larger than the visible range ($u$, $v \in [-3, 3]$) [Fig. 8(*a*)] for the arrangement in Fig. 6(*a*) shows that (*i*) the *GL*s only appear along the directions $(u_{bc}, v_{bc})$, $b, c \in \mathbb{N}$, $b \times c \neq 0$ (red circles in Fig. 8) as predicted by (25), (*ii*) despite the massive thinning [i.e., $\tau \approx 0.5$ - Fig. 6(*a*)] and the isophoric nature of the excitations, the *DS* properties guarantee that no grating/major sidelobes appear beyond those of the corresponding (i.e., displaced over the same array lattice) fully-populated layout [Fig. 8(*b*)]. This property, which is confirmed also varying the lattice geometry [e.g., $\mathbf{d}_1 = \frac{\lambda}{2}\widehat{\mathbf{x}}$, $\mathbf{d}_2 = \left(\frac{3\lambda}{10}\widehat{\mathbf{x}} + \frac{\lambda}{2}\widehat{\mathbf{y}}\right)$ - Fig. 8(*d*); $\mathbf{d}_1 = \frac{\lambda}{2}\widehat{\mathbf{x}}$, $\mathbf{d}_2 = \left(\frac{\lambda}{2}\widehat{\mathbf{x}} + \frac{\lambda}{2}\widehat{\mathbf{y}}\right)$ - Fig. 8(*g*)], also proves that *DS*-based massively-thinned isophoric arrays have the same field-of-view (i.e., the maximum one unless dealing with non-regular distributions) of the corresponding non-thinned arrangements. Otherwise, other grating lobes, beyond those due to the lattice geometry (25) [e.g., Fig. 8(*c*) vs. Fig. 8(*b*)], arise in alternative thinned architectures such as, for instance, randomly thinned arrays [Fig. 8(*c*), Fig. 8(*f*), and Fig. 8(*i*)].

### C. *Analytical Isophoric Thinning*

Let us now analyze the potentialities and the features of (48) as a design tool for isophoric thinned arrangements complying with user-defined constraints on the figures of merit of the radiation pattern of main interest. Towards this end, the first test case consists of the following requirements: sidelobe level threshold $SLL^T = -23.0$ [dB], minimum directivity $D^T = 29.0$ [dB], a pattern value equal to $\mathcal{P}^T = \mathcal{P}(u_0, v_0) - 30.0$ [dB] at $\left(u^T, v^T\right) = \left(5.3 \times 10^{-1}, 4.5 \times 10^{-2}\right)$, and a half-power beamwidth smaller than $BW^T = 6.00$ [deg] (Tab. II).

According to the design procedure presented in Sect. IV-B, the first step (*Step 1*) is the selection of the $(P \times Q, H, \gamma)$-*DS* $\Xi$ so that $SLL_{SUP} \leq SLL^T$ (38). By searching among the available *DS* sequences in [28][29], the $(1023, 511, 255)$-*DS* taken from [28] can be considered as a potential candidate sequence since, by simple substitution, $SLL_{SUP} = \frac{\epsilon(H-\gamma)[0.5+1.5\log_{10}(P\times Q)]}{\gamma\times(P\times Q-1)+H} = -23.1$ [dB] (Tab. II).

The *Step 2* is devoted to choose, given $\Xi$, the array lattice to fit the requirements on the pattern value along the direction $\left(u^T, v^T\right)$ also avoiding the occurrence of *GL*s [$\sqrt{u_{bc}^2 + v_{bc}^2} > 1$, $\forall\ b, c \in \{-1, 0, 1\}$ (25)]. Preliminarily, it is needed to check through (43) whether $\Xi$ is admissible to this purpose. In this case, such a condition is verified since $(H - \gamma) \times \mathcal{P}_{el}\left(u^T, v^T\right) = 256$ is smaller than $\mathcal{P}^T = \mathcal{P}_{el}(u_0, v_0) \times [\gamma \times (P \times Q - 1) + H] - 30$ [dB] $= 266.221$. Then, $\mathbf{d}_1$ and $\mathbf{d}_2$ are determined so that $\left(u_0 + \lambda\frac{\frac{md_{2y}}{P} - \frac{nd_{1y}}{Q}}{d_{1x}d_{2y} - d_{2x}d_{1y}},\ v_0 + \lambda\frac{\frac{nd_{1x}}{Q} - \frac{md_{2x}}{P}}{d_{1x}d_{2y} - d_{2x}d_{1y}}\right) = \left(u^T, v^T\right)$. This means defining the integers $m$ and $n$ as well as ($d_{1x}$, $d_{1y}$, $d_{2x}$, $d_{2y}$) from a single equation (44), that is having a multiplicity of possible solutions. Therefore, an admissible choice is that with $m = 8$, $n = 3$, $\mathbf{d}_1 = 0.47\lambda\widehat{\mathbf{x}} + 0.21\lambda\widehat{\mathbf{y}}$, $\mathbf{d}_2 = 0.12\lambda\widehat{\mathbf{x}} + 0.61\lambda\widehat{\mathbf{y}}$ (Tab. II).

Successively (*Step 3*), the compliance of the trial set $\{\underline{\Xi}$, $m = 8$, $n = 3$, $\mathbf{d}_1$, $\mathbf{d}_2\}$ at hand with the remaining bounds (41)(47) is assessed. By substituting $P$, $Q$, $H$, $\gamma$, $\mathbf{d}_1$, and $\mathbf{d}_2$ in (41) and (47), it turns out that $D_{INF} = 29.1$ [dB] is greater than $D^T$ (Tab. II) and $BW_{SUP} = 5.65$ [deg] is smaller than $BW^T$ (Tab. II) as required by the project guidelines. The last step of the synthesis process (*Step 4*) is then concerned with the definition of the array layout from the sequence $\underline{\alpha}^{\left(\sigma^{opt}\right)}$, that is the optimal cyclic-shift of the $(P \times Q, H, \gamma)$-*DS* $\underline{\Xi}$. As it can be inferred from Fig. 9, where the plots of $SLL^{(\sigma)}$ [Fig. 9(*a*)], $D^{(\sigma)}$ [Fig. 9(*b*)], and $BW^{(\sigma)}$[Fig. 10(*c*)] versus the shift index $\sigma$ are shown, it turns out that the optimal shift is $\sigma^{opt} = 874$. For completeness, the arising array layout [Fig. 10(*a*)] and its corresponding power pattern [Fig. 10(*b*)] as well as the associated quality indexes (Tab. II) are given. Once again it is worth pointing out that these results, which faithfully match the design guidelines/constraints, are yielded by just using isophorically-excited elements displaced on a regular lattice without recurring to excitation tapering and/or sparse element arrangements nor exploiting any iterative optimization.

Table II
*Analytical Isophoric Thinning* - REQUIREMENTS, *DS* PERFORMANCE BOUNDS, AND SYNTHESIZED *DS* LAYOUTS WITH PERFORMANCE INDEXES.

| | *Figure* | 10 | 11(*b*) | 11(*c*) |
|---|---|---|---|---|
| *Req.* | $SLL^T$ [dB] | $-23.0$ | $-18.0$ | – |
| | $D^T$ [dB] | 29.0 | 22.0 | – |
| | $\mathcal{P}^T$ [dB] | $-30.0$ | $-25.0$ | – |
| | $BW^T$ [deg] | 6.00 | 12.00 | – |
| *Aperture* | $P$ | 31 | 16 | 16 |
| | $Q$ | 33 | 16 | 16 |
| *Lattice* | $d_{1x}$ [λ] | 0.47 | 0.50 | 0.50 |
| | $d_{1y}$ [λ] | 0.21 | 0.00 | 0.00 |
| | $d_{2x}$ [λ] | 0.12 | 0.30 | 0.30 |
| | $d_{2y}$ [λ] | 0.61 | 0.50 | 0.50 |
| *Analytical Bounds* | $SLL_{SUP}$ [dB] | $-23.1$ | $-18.5$ | – |
| | $D_{INF}$ [dB] | 29.1 | 22.4 | – |
| | $\mathcal{P}\left(u^T, v^T\right)$ [dB] | $-30.0$ | $-25.2$ | – |
| | $BW_{SUP}$ [deg] | 5.65 | 11.78 | – |
| *Results* | $SLL$ [dB] | $-23.8$ | $-19.9$ | $-20.2$ |
| | $D$ [dB] | 30.9 | 26.3 | 27.4 |
| | $\mathcal{P}\left(u^T, v^T\right)$ [dB] | $-30.0$ | $-25.2$ | $-27.1$ |
| | $BW^{\max}$ [deg] | 4.55 | 8.84 | 8.82 |

The last experiment is aimed at assessing the effectiveness of (48) when arrays with *non-isotropic* [i.e., $\mathcal{E}(u,v) \neq 1.0$] radiators have to be synthesized by also evaluating the arising analytically-designed layouts by means of *full-wave* numerical simulations. As an illustrative example, the synthesis of an isophoric thinned layout featuring half-wavelength $y$-directed dipoles affording a radiation pattern with $SLL \leq SLL^T$ ($SLL^T = -18.0$ [dB]), $D \geq D^T$ ($D^T = 22.0$ [dB]), $BW^{\max} \leq BW^T$ ($BW^T = 12.0$ [deg]), and having a value smaller than $\mathcal{P}^T = \mathcal{P}(u_0, v_0) - 25.0$ [dB] along the direction $\left(u^T, v^T\right) = \left(-3.8 \times 10^{-1}, -2.5 \times 10^{-2}\right)$. By modeling the element factor with the standard *cosine* approximation (i.e., $\mathcal{E}(u,v) \approx \sqrt{1-u^2}$) [31][32] within the design process, the Hadamard *DS* with descriptors $P=Q=16$, $H=136$, and $\gamma=72$ [29] is chosen (*Step 1*) being (38) $SLL_{SUP} = -18.5$ [dB] (Tab. II). As for the lattice grid, the equation (44) is inverted to find one of its admissible solutions (*Step 2*). For instance, the trial set $\{m=-3,\ n=-2,\ \mathbf{d}_1 = 0.5\lambda\widehat{\mathbf{x}},\ \mathbf{d}_2 = 0.3\lambda\widehat{\mathbf{x}} + 0.5\lambda\widehat{\mathbf{y}}\}$ (Tab. II) has been selected since $\mathcal{P}\left(u^T, v^T\right) = -25.2$ [dB] (Tab. II) as theoretically expected (44). Being (41) and (47) satisfied (*Step 3*) since $D_{INF} = 22.4$ [dB] (Tab. II) and $BW_{SUP} = 11.78$ [deg] (Tab. II), the generator Hadamard *DS* sequence $\Xi$ is kept as viable, even though not the best achievable through cyclic shift (*Step 4*), solution of the problem at hand.

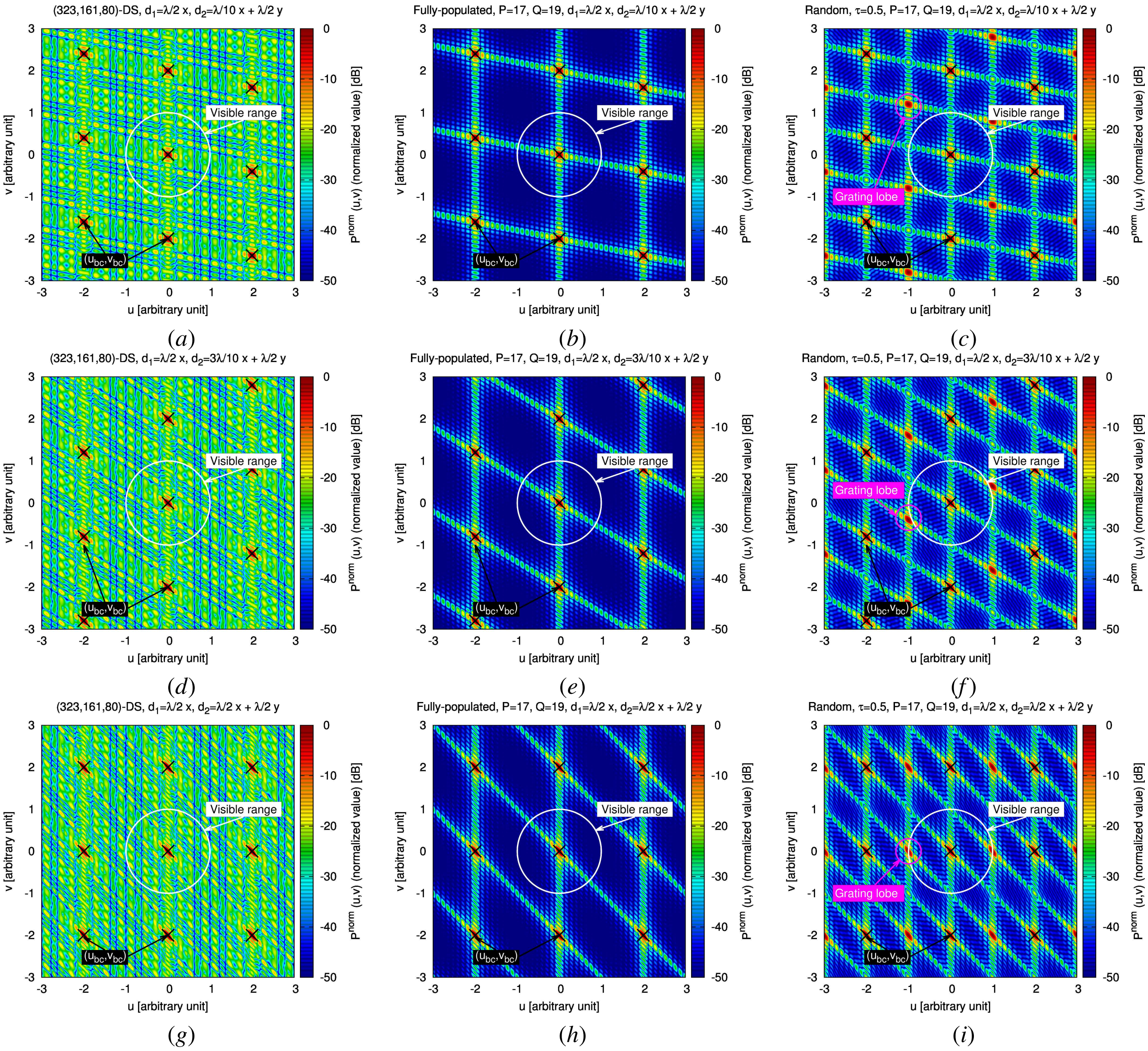

Figure 8. *Performance Bounds* ($P=17$, $Q=19$, $H=161$, $\gamma=80$, $\mathcal{E}(u,v)=1.0$, $\theta_0=\varphi_0=0.0$ [deg]) - Normalized power pattern $\mathcal{P}^{norm}(u,v)$ and grating lobe positions $(u_{bc}, v_{bc})$ (25) in the range $u,\, v \in [-3, 3]$ when $(a)(b)(c)$ $\mathbf{d}_1=\frac{\lambda}{2}\widehat{\mathbf{x}}$, $\mathbf{d}_2=\left(\frac{\lambda}{10}\widehat{\mathbf{x}}+\frac{\lambda}{2}\widehat{\mathbf{y}}\right)$, $(d)(e)(f)$ $\mathbf{d}_1=\frac{\lambda}{2}\widehat{\mathbf{x}}$, $\mathbf{d}_2=\left(\frac{3\lambda}{10}\widehat{\mathbf{x}}+\frac{\lambda}{2}\widehat{\mathbf{y}}\right)$, and $(g)(h)(i)$ $\mathbf{d}_1=\frac{\lambda}{2}\widehat{\mathbf{x}}$, $\mathbf{d}_2=\left(\frac{\lambda}{2}\widehat{\mathbf{x}}+\frac{\lambda}{2}\widehat{\mathbf{y}}\right)$ for $(a)(d)(g)$ the *DS*-based layout and its $(b)(e)(h)$ corresponding fully-populated and $(c)(f)(i)$ random architectures.

The power pattern [Fig. 11($b$)] and the figures of merit (Tab. II) of the analytically-designed arrangement [Fig. 11($a$)] are then compared to those obtained from the model of the same architecture with *FEKO* [Fig. 11($c$), Tab. II]. From the comparison, one can infer that ($i$) the proposed design technique applies seamlessly to non-isotropic radiators yielding low and controlled sidelobes [Fig. 11($b$)] as confirmed by the full-wave simulation in [Fig. 11($c$)]; ($ii$) the figures of merit of the "ideal" layout [i.e., $SLL = -19.9$ [dB]; $D = 26.3$ [dB]; $\mathcal{P}\left(u^T, v^T\right) = -25.2$ [dB]; $BW^{\max} = 8.84$ [deg] - Fig. 11($b$) and Tab. II] and of the "full-wave" modeled one [i.e.,

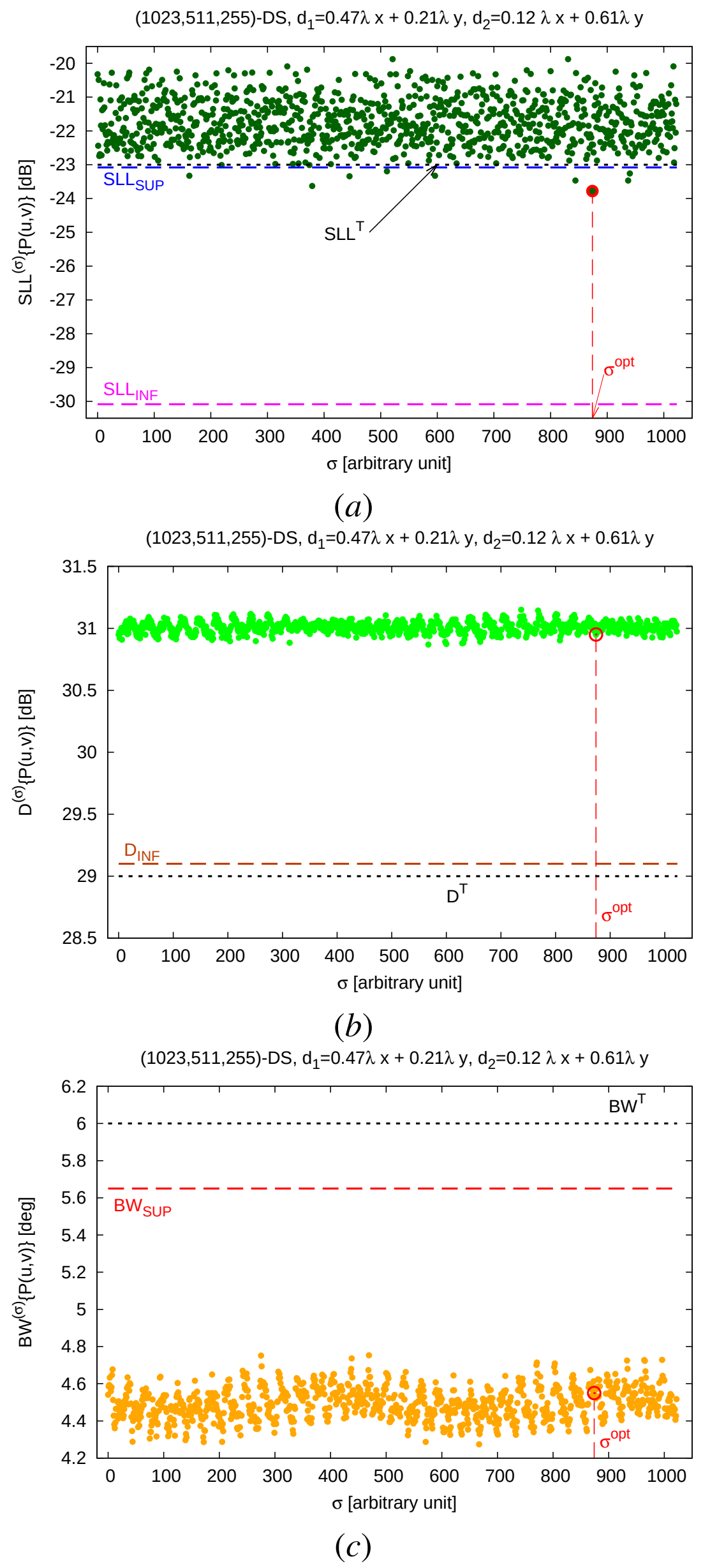

Figure 9. *Analytical Isophoric Thinning* ($SLL^T = -23.0$ [dB], $D^T = 29.0$ [dB], $\mathcal{P}^T = -30.0$ [dB], $u^T = 5.3{\times}10^{-1}$, $v^T = 4.5{\times}10^{-2}$, $BW^T = 6.0$ [deg], $\mathcal{E}\left(u, v\right) = 1.0$) - Behaviour of ($a$) the $SLL$ ($b$) the $D$, and ($c$) the $BW^{\max}$ of isophoric arrays derived from the *DS* sequence $\underline{\alpha}^{(\sigma)}$ versus the shift index $\sigma$.

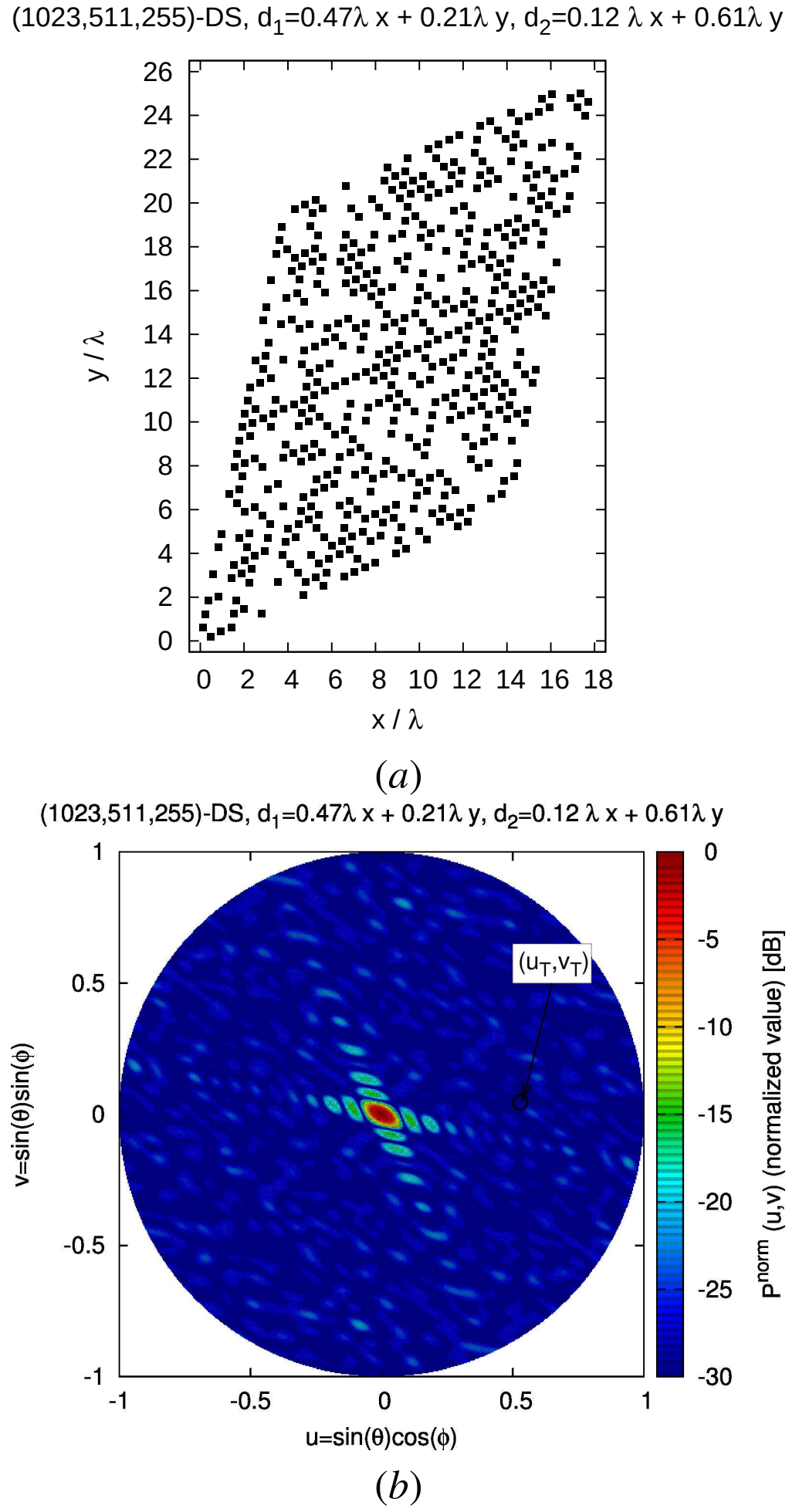

Figure 10. *Analytical Isophoric Thinning* ($SLL^T = -23.0$ [dB], $D^T = 29.0$ [dB], $\mathcal{P}^T = -30.0$ [dB], $u^T = 5.3 \times 10^{-1}$, $v^T = 4.5 \times 10^{-2}$, $BW^T = 6.0$ [deg], $\mathcal{E}\left(u, v\right) = 1.0$) - Plot of ($a$) the $(1023, 511, 255)$-*DS* array layout ($\sigma^{opt} = 874$) and ($b$) the corresponding radiated power pattern $\mathcal{P}^{norm}\left(u, v\right)$.

$SLL = -20.2$ [dB]; $D = 27.4$ [dB]; $\mathcal{P}\left(u^T, v^T\right) = -27.1$ [dB]; $BW^{\max} = 8.82$ [deg] - Fig. 11($b$) and Tab. II] are both fully compliant with the design objectives. This latter outcome is a further proof that the proposed approach, despite its fully-analytic nature, can be profitably employed for the design of isophoric thinned arrangements with *a-priori* performance predictions even when realistic structures (i.e., arrangements characterized by mutual coupling effects and deviations from the ideal case) are taken into account.

## VI. Conclusions and Final Remarks

An innovative method for the synthesis of isophoric thinned arrays with controlled pattern features has been proposed as a final product of a more general theoretical framework. At first, the analytic expression for the prediction of the power pattern samples of arbitrary array lattices and excitations has been derived. Then, closed-form bounds have been deducted for the *a-priori* estimation of the radiation properties of weighting sequences with known autocorrelation properties. Representative results from a set of numerical experiments have been reported and discussed to give some insights on the reliability and the accuracy of the analytic performance bounds as well as to validate the arising design procedure and synthesis guidelines also through comparisons with full-wave simulations.

The main methodological advancements of this paper with respect to the state-of-the-art can be then summarized as follows: ($a$) the derivation of the general theory that relates the array power pattern and the autocorrelation of its excitations for arbitrary lattices and array weights (Sect. II), ($b$) the deduction of closed-form expressions for the performance bounds of isophoric thinned arrays (Sect. IV-A), ($c$) the definition of a synthesis procedure with analytic design equations for the synthesis of *DS*-based isophoric thinned arrays complying with pattern features user-defined and, unlike current analytical thinning techniques [16][17], not only on the sidelobe level, (*d*) the generalization of the *DS*-based thinning theory comprising existing analytic design techniques [16] as a particular case (Sect. IV).

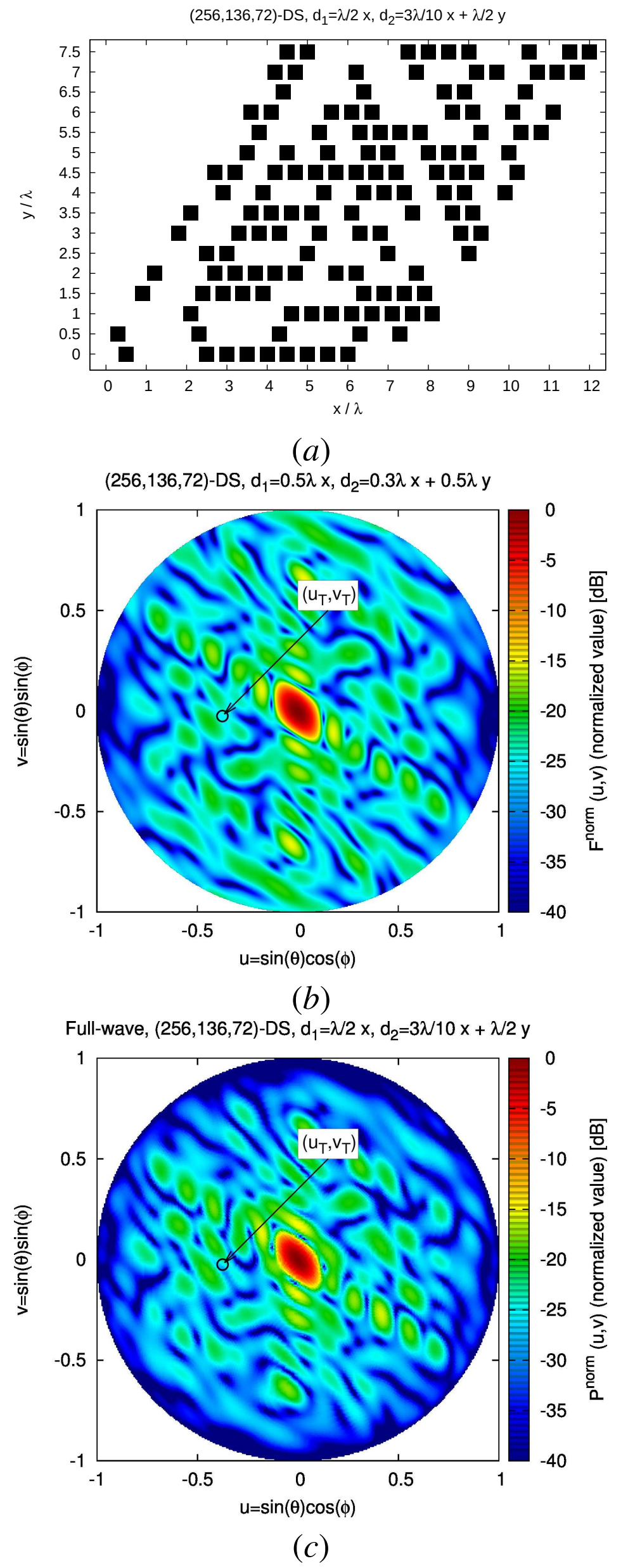

Figure 11. *Analytical Isophoric Thinning* ($SLL^T = -18.0$ [dB], $D^T = 22.0$ [dB], $\mathcal{P}^T = -25.0$ [dB], $u^T = -3.7 \times 10^{-1}$, $v^T = -2.5 \times 10^{-2}$, $BW^T = 12.0$ [deg], $\frac{\lambda}{2}$-dipole radiators) - Plot of (*a*) the synthesized $(256, 136, 72)$-*DS* array layout and the corresponding radiated power pattern $\mathcal{P}^{norm}(u, v)$ computed with (*b*) (1) and (*c*) the *FEKO* full-wave *MoM*-based solver.

From the results of the numerical analysis, the following guidelines and design principles have been inferred. More specifically, (*a*) *for any excitation sequence* $\underline{\alpha}$ *and whatever two-dimensional array lattice*:

- the samples of the power pattern along known (8) directions can be analytically predicted (15) without requiring the knowledge of the excitation sequence, $\underline{\alpha}$, but only the values of its autocorrelation function $\underline{a}$;
- the power pattern in the whole $(u, v)$-space can be faithfully determined (16) from the knowledge of the excitation sequence, $\underline{\alpha}$, starting from the pattern samples (15), the computation of the *DFT* of the excitation sequence, and the exploitation of the interpolation function (12);
- the power pattern in the whole $(u, v)$-space can be approximated (17) from the knowledge of the autocorrelation function $\underline{a}$ and the exploitation of the interpolation function (12);

(*b*) *for binary DS excitation sequences* ($\underline{\alpha} \equiv \underline{\Xi}$) *and whatever two-dimensional array lattice*:

- the power pattern samples of a sequence can be *a-priori* determined (23)(24) by only using the descriptors (i.e., $P$, $Q$, $H$, and $\gamma$) of the *DS* sequence;
- the position of grating lobes as well as the field-of-view can be *a-priori* computed (25) from the knowledge of the array lattice descriptors (i.e., $d_{1x}$, $d_{1y}$, $d_{2x}$, and $d_{2y}$);
- analytic bounds for the sidelobe level (38), the directivity (41), and the half-power beamwidth (47) can be reliably predicted by just knowing the descriptors of the generating sequence $\underline{\Xi}$ (i.e., $P$, $Q$, $H$, and $\gamma$).

Future works, beyond the scope of the current paper, will be aimed at exploiting the theoretical framework in (Sect. II) to define innovative synthesis methods where the problem of designing an array with user-defined pattern features is recast to the synthesis of a/different suitable autocorrelation function/s. Moreover, the extension to more complex geometries (e.g., cylindrical/conformal architectures) is at a very preliminary stage since currently under theoretical investigation.

## Acknowledgements

A particular thank to E. Vico for her never-ending support, guidance, and help.